\documentclass[12pt,preprint]{aastex}
\bibpunct[, ]{(}{)}{;}{a}{,}{,}
\newcommand{\be}{\begin{equation}}
\newcommand{\ee}{\end{equation}}
\begin{document}
\title{Magnetic Field Structure and
Stochastic Reconnection in a Partially Ionized Gas}

\author{ A. Lazarian}

\affil{Dept. of Astronomy, University of Wisconsin, 475 N Charter Street,
Madison WI 53706}

\email{lazarian@astro.wisc.edu}

\author{Ethan T. Vishniac}

\affil{Department of Physics and Astronomy,
Johns Hopkins University,
Baltimore MD 21218}
\email{ethan@pha.jhu.edu}

\and

\author{ Jungyeon Cho}

\affil{Dept. of Astronomy, University of Wisconsin, 475 N Charter Street,
Madison WI 53706}

\email{cho@astro.wisc.edu}

\begin{abstract}
We consider stochastic reconnection in a magnetized, partially ionized
medium.  Stochastic reconnection is a generic effect, due to field
line wandering, in which the speed of reconnection is determined by 
the ability of ejected plasma to diffuse away from the current sheet
along magnetic field lines, rather than by the details of current 
sheet structure.   As in earlier work, in which we dealt with a fully
ionized plasma, we consider the limit of weak stochasticity, so that
the mean magnetic field energy density is greater than either the 
turbulent kinetic energy density or the energy density associated with
the fluctuating component of the field.  For specificity, we consider
field line stochasticity generated through a turbulent cascade, which
leads us to consider the effect of neutral drag on the turbulent 
cascade of energy.  In a collisionless plasma, neutral particle 
viscosity and ion-neutral drag will damp mid-scale turbulent motions, 
but the power spectrum of the magnetic perturbations extends below 
the viscous cutoff scale. We give a simple physical picture of the 
magnetic field structure below this cutoff, consistent with numerical
experiments.  We provide arguments for the re\"emergence of the 
turbulent cascade well below the viscous cut-off scale and derive 
estimates for field line diffusion on all scales.  We note 
that this explains the persistence of a single power law form for the 
turbulent power spectrum of the interstellar medium, from scales of 
tens of parsecs down to thousands of kilometers.  We find that under 
typical conditions in the ISM stochastic reconnection speeds are 
reduced by the presence of neutrals, but by no more than an order of
magnitude.  However, neutral drag implies a steep dependence on the
Mach number of the turbulence.  In the dense cores of $H_2$ regions 
the reconnection speed is probably determined by 
tearing mode instabilities.
\end{abstract}

\keywords{Magnetic fields; Galaxies: magnetic fields,
ISM: molecular clouds, magnetic fields; Stars: formation }

\section{Introduction}

One of the fundamental properties of astrophysical magnetic fields
is their ability to change topology via reconnection
\citep[see][]{PF00}.  It is impossible to understand the
origin and evolution of large scale magnetic fields without
understanding the mobility of magnetic field lines. In a typical
astrophysical plasma, resistivity is very small
and flux freezing, which follows from assuming zero
resistivity, should be an excellent guide to the
motion of magnetic fields. In spite of this, fast
magnetic dynamo theory
\citep[see][]{P79,M78,KR80} invokes a
constantly changing magnetic field topology and motions during
which field lines cross each other\footnote{While standard mean-field
dynamo theory has been severely criticized on theoretical
and numerical grounds \citep{VC92,GD94,GD96,CH96,HCK96,B01},
an alternative version can be
formulated which evades these criticisms \citep[][]{VC01} and which
assumes fast reconnection only on two dimensional surfaces.}.
This assumption is supported
by observations of the solar magnetic field
\citep[see][and references contained therein]{D96,IIAW97}
which are difficult to explain unless flux
freezing is routinely violated on time
scales short compared to resistive time scales, at least within
thin current sheets.

The Sweet-Parker model of reconnection is the simplest and most robust
\citep{P57, S58}.  In this model, reconnection takes place within
a thin current sheet, which separates two large volumes containing
uniform, and very different, magnetic fields.  The resulting reconnection speed
is less than the Alfv\'en speed
by the square root of the Lundquist number
$\sim {\cal R}_L^{-1/2}=(\eta/V_A L)^{1/2}$, where
$\eta$ is the resistivity, $V_A$ is the Alfv\'en speed, and $L$
is the length of the current sheet, assumed to be determined
by the large scale geometry of the problem.  Under typical astrophysical
conditions this is very slow (e.g. for the Galaxy as a whole
${\cal R}_L\sim 10^{20}$).  This reconnection
speed is set by a geometrical constraint. Indeed, 
plasma tied to the reconnecting magnetic field lines must be ejected
from the ends of the narrow current sheet. The disparity of scales, one
of which is macroscopic/astrophysical, while the other is microscopic, 
i.e. determined by ohmic diffusion, makes the reconnection slow.

This evident shortcoming of the Sweet-Parker reconnection
scheme has stimulated interest in
alternative models that allow fast reconnection. Although the
literature on magnetic reconnection is extensive
\citep[e.g.][and references therein]{PF00} it does not
successfully address this question. Models
that invoke an X-point reconnection geometry \citep{P64} have
been
shown to be unstable for sufficiently high $R_L$ \citep[see][]{B96},
while anomalous resistivity fails to provide rapid reconnection
under most astrophysical conditions \citep[see][]{P79}.
A general review of astrophysical magnetic reconnection
theory can be found in \citet{BMW03}. 

A notable exception to this discouraging state of affairs is 
the recent work on fast collisionless
reconnection
\citep[see also the discussion by Bhattacharjee, Ma and Wang 2001]
{BSD97,SDD98,SD98}.
This
work indicates that under some circumstances a kind of standing
whistler mode can stabilize an X-point reconnection region.  However,
these studies have not demonstrated the possibility of fast reconnection
for generic field geometries. They assume that there are no
bulk forces acting to produce a large scale current
sheet, and that the magnetized regions are convex, which minimizes
the energy required to spread the field lines.  In addition, 
this mechanism requires a collisionless
environment, where the electron mean free path is less than the current sheet
thickness \citep{TYJKC03, JYHK98}.
Unfortunately, in the laboratory the current sheet thickness
is comparable to the ion Larmor radius and
it is unclear how to generalize this criterion
to the interstellar medium, where the Sweet-Parker current sheet thickness
is typically much greater than the Larmor radius.  {\it If} we require that
the ion skin depth, the characteristic scale of the standing Whistler mode,  
be greater than the current sheet thickness, then we
have a criterion which is rarely satisfied in the interstellar medium.
In this paper we will concentrate on a mechanism that acts in turbulent media
and when it works,
produces rapid reconnection under a broad range of field geometries,
without regard to the particle collision rate.  We will defer all discussion
of the relationship between stochastic reconnection and small scale
collisionless effects to a later paper.


In a previous paper \citep[henceforth LV99]{lv99}
we discussed `stochastic reconnection', a process which is
similar to Sweet-Parker reconnection, except that stochastic
wandering of the field lines produces a broad outflow region.
The properties of the outflow region are insensitive to the width of the
current sheet (and the value of ${\cal R}_L$), but depend on the
level of field line stochasticity.  In a sufficiently noisy environment
the reconnection speed becomes a large fraction of the Alfv\'en speed.
In an {\it extremely} quiet environment the field lines do not
enter or leave the current sheet over its entire length and we
recover the Sweet-Parker reconnection model.
In LV99 we dealt with an inviscid and totally ionized fluid.
\footnote{In this paper we shall show that
this approximation is valid for partially ionized collisionless plasma
up to a certain percentage of neutrals, and for collisional gases with
resistivity larger than viscosity.}

The notion that magnetic field stochasticity might affect 
current sheet structures is not unprecedented.  In earlier
work \citet{S70} showed that in collisionless plasmas the
electron collision time should be replaced with the time a
typical electron is retained in the current sheet.  Also
\citet{JM84} proposed that current diffusivity should be modified
to include diffusion of electrons across the mean field due
to small scale stochasticity.  These effects will usually be small
compared to effect of a broad outflow zone containing both
plasma and ejected shared magnetic flux.  Moreover,
while both of these effects
will affect reconnection rates, they are not sufficient to 
produce reconnection speeds comparable to the Alfv\'en speed
in most astrophysical environments.    

It is important to distinguish between stochastic reconnection,
as discussed in LV99, and the more conventional notion of
turbulent reconnection.  The latter usually involves substituting a
turbulent diffusivity for the resistivity, and involves
a degree of small-scale mixing which is forbidden
on energetic grounds \citep[see \S 2 in][and the references
contained therein]{P92}.  A common variation of this hypothesis
is that instabilities
in the current sheet will produce a hugely broadened current sheet
and a large effective resistivity within it \citep[for a review see][]
{B00}.  On the
contrary, the former is largely a topological effect, using conventional
estimates of resistivity, and the only strong mixing associated
with it has to do with the polarization of field lines crossing
the current sheet.  (That is, one expects sharp gradients, in the
current sheet, in the component of the magnetic lines perpendicular
to the current sheet.)
In this sense, stochastic reconnection belongs
to the class of models which try to explain fast reconnection by
appealing to a current sheet geometry which is `natural' in some
sense, but evades the limits set by the Sweet-Parker model.  (The
bulk of the discussion in \citet{PF00} is centered on laminar
three dimensional field configurations which can lead to similarly
rapid reconnection speeds.)

In a recent paper Kim \& Diamond (2001) addressed the problem of stochastic
reconnection by calculating the turbulent diffusion rate for magnetic flux
inside a current sheet.  They obtained similar turbulent 
diffusion rates for both two dimensional and reduced three 
dimensional MHD. In both cases the presence of turbulence had a
negligible effect on the flux transport.  The authors pointed out
that this would prevent the anomalous transport of magnetic flux within the current
sheet and concluded that both 2D and 3D stochastic reconnection proceed
at the Sweet-Parker rate even if individual small scale reconnection
events happen quickly. If true this would be not only rule out 
the LV99 reconnection scheme, but also any other fast reconnection scheme.
In general astrophysical plasmas are turbulent and if the 
enhancement of the local reconnection speed, e.g. due to collisionless 
effects (see Drake et al. 2001), is irrelevant then reconnection must always 
be slow.

However turbulent diffusion rates within the current sheet are irrelevant for the 
process of stochastic reconnection or, for broadly similar reasons, fast 
collisionless reconnection.
The basic claim in LV99 is that realistic magnetic field topologies allow multiple
connections between the current sheet and the exterior environment, which would
persist even if the stochastic magnetic field lines were stationary ("frozen in 
time") before reconnection.  This leads to global outflow constraints which
are weak and do not depend on the properties of the current sheet.  In particular,
the analysis in LV99 assumed that the current sheet thickness is
determined purely by ohmic dissipation and that turbulent diffusion of the magnetic
field is negligible inside, and outside, the current sheet.  The major uncertainty 
in this model is the behavior of the reconnected flux elements, which are nearly 
perpendicular to the current sheet and must undergo multiple reconnections before 
being ejected.  We note also that models of collisionless reconnection also evade
the objection posed by Kim and Diamond topologically, that is, by stabilizing 
an X-point reconnection topology, and opening up the rest of the current sheet.

"Hyper-resistivity" \citep{S85,BH86,HB87} is a more subtle attempt to derive fast reconnection
from turbulence within the context of mean-field resistive MHD.  The form of the parallel
electric field can be derived from magnetic helicity conservation.  Integrating
by parts one obtains a term which looks like an effective resistivity
proportional to the magnetic helicity current.  There are several assumptions
implicit in this derivation, but the most important problem is that by
adopting a mean-field approximation one is already assuming some sort of
small-scale smearing effect, equivalent to fast reconnection.  \citet{S88}
partially circumvented this problem by examining the effect of tearing
mode instabilities within current sheets.  However, the resulting reconnection
speed enhancement is roughly what one would expect based simply on the
broadening of the current sheets due to internal mixing.  This effect
does not allow us to evade the constraints on the global
plasma flow that lead to slow reconnection speeds, a point which
has been demonstrated numerically \citep{ML85}
and analytically (LV99).  Nevertheless, we show in \S 4 that this effect may be 
important in the densest and coldest parts of the ISM.

A partially ionized plasma fills a substantial volume within
our galaxy and the earlier stages of star formation take
place in a largely neutral medium. This motivates our study
of the effect of neutrals on reconnection.
The role of ion-neutral collisions is not trivial. On one hand, they
may truncate the turbulent cascade, reducing the small scale 
stochasticity and decreasing the reconnection
speed. On the other hand, the ability of neutrals to diffuse perpendicular
to magnetic field lines allows for a broader particle outflow 
and enhances reconnection rates.

Reconnection in partially ionized gases has been already studied by
various authors \citep{NMA92,ZB97}.  In a 
recent study \citep[henceforth VL99]{VL99}
we studied the diffusion of neutrals away from the reconnection zone
assuming anti-parallel magnetic field lines \citep[see also][]{HZ03a}
The ambipolar reconnection rates obtained in VL99, although large
compared with the Sweet-Parker model, are insufficient either
for fast dynamo models or for the ejection of magnetic flux prior
to star formation.  In fact, the increase in the reconnection speed
stemmed entirely from the
compression of ions in the current sheet, with the consequent enhancement of
both recombination and ohmic dissipation.  This effect is small
unless the reconnecting magnetic field lines are almost exactly
anti-parallel \citep{VL99, 
HZ03b}.  Any dynamically significant shared field component
will prevent noticeable plasma compression in the current
sheet, and lead to speeds practically indistinguishable from the standard
Sweet-Parker result.  Since generic reconnection regions will have
a shared field component of the same order as the reversing component,
the implication is that reconnection and ambipolar diffusion do not
change reconnection speed estimates significantly.

None of this work on reconnection in partially ionized
plasmas includes the effects of
stochasticity. We expect that in the presence of turbulence,
reconnection rates will be substantially enhanced, as they are in
completely ionized plasmas. To generalize the concept of stochastic
reconnection to partially ionized plasmas we need a model for the
small scale stochasticity of a turbulent magnetic field in a partially
ionized plasma. In this paper we will begin by considering this
problem, and then apply our results to the reconnection speed.

Following LV99, we consider reconnection in the presence of a weakly
stochastic
magnetic field.  Except for the presence of noise, we imagine a
reconnection event exactly like a generic Sweet-Parker reconnection
event.  Two volumes with average magnetic fields that are of comparable
strength, but differing directions are in contact over a surface of
length $L$.  Due to the stochastic nature of the fields, field lines come
into contact over many small patches (see Fig.~1).
For each individual patch the Sweet-Parker reconnection model
should be applicable (or at least constitute a minimal reconnection speed).
The enhancement of reconnection rates follows from two effects.
First, since individual field lines wander out of the narrow current
sheet relatively easily, the longitudinal patch size is much smaller
than the overall size of the system.
This reduces the effective value of ${\cal R}_L$ and raises
the local reconnection speed. Second, whereas in the Sweet-Parker
scheme magnetic field lines reconnect sequentially, in the
presence of field line wandering field
lines many independent patches are brought into direct contact
and can reconnect simultaneously.  
As a consequence, the rate of reconnection of the magnetic flux 
is increased by a large factor, whose exact value depends on the level of
noise in the system.

In LV99 the truncation of the turbulent cascade was assumed to be
due to resistivity.  Consequently, the smallest scale of field line
wandering decreases as resistivity decreases, and the number of
independent patches in contact within the reconnection zone increases.
{}From this we concluded that for an idealized inviscid
fluid the reconnection rate does not depend on
fluid resistivity. It does depend on the level
of magnetic field stochasticity.  In the specific case where
the field line stochasticity is caused by a turbulent cascade,
it depends on the amplitude of the turbulence.  However, there
is no necessary connection between turbulent motions and the
reconnection speed.  In particular, the rate of turbulent transport
of mean magnetic flux is assumed to be negligible in this model,
and even the complete absence of turbulent diffusion in the
current sheet would not reduce the stochastic reconnection rate
\citep[cf.][]{KD01}.

To quantify stochastic reconnection we have to use a particular
description of turbulence.
Motions in a magnetized medium can be expanded into incompressible (Alfv\'en)
and compressible (fast and slow) modes. There are theoretical arguments
\citep{gs95,LG01} \citep[see][for a review]{CLV03a} 
suggesting that the nonlinear cascade of power for these
modes proceeds separately, although not entirely independently.
Simulations in \citet{CL02} support this idea
and show that the  \citet[henceforth GS95]{gs95}
scaling is applicable to Alfvenic part of the MHD cascade.
For our purposes the scalings of slow and fast modes \citep{CL02}, \citep{CL03}
are less important since they are subjected to collisionless
damping\footnote{Whether or not those damped modes are important depends
on the process studied. For instance, \citet{YL02} show that
for scattering of cosmic rays the residual small amplitude
fast modes are much more efficient than the Alfv\'en modes. This,
however, is not true for the field wandering that we consider in this
paper.}

Here we assume that the
GS95 model describes incompressible turbulence above the ambipolar
damping scale\footnote{We note, however,
that our qualitative conclusions for reconnection rates should be
valid for other models of MHD turbulence (see LV99) as long as they are in
rough agreement with observational constraints and numerical simulations.}.
To describe MHD turbulence below the scale of
viscous damping we present a new model of magnetic field structure in this
regime.
This model is in rough agreement with numerical simulations by
\citet[henceforth CLV02b]{CLV02b}.
%

%

In \S 2 of this paper we will consider the effect of a large
neutral fraction on a strongly turbulent cascade in a magnetized
plasma.  In \S 3 we apply this to the problem of reconnection
in partially ionized plasmas. In \S 4 we apply this work to various
phases in the ISM.  Finally, \S 5 contains our basic conclusions.

\section{Magnetohydrodynamic Turbulence in a Partially Ionized Plasma}

In this section we will consider the effect of neutral particles on the turbulent cascade 
in the ISM.  We begin by briefly reviewing the nature of the strong turbulent
MHD cascade and the dynamical influence of neutral particles.  In \S 2.2 we 
describe the cascade when viscous damping, due to neutral particles, is
strong, but the one fluid approximation remains valid.  In \S 2.3 we consider
the uncoupled regime, covering scales where the neutral particles exert
a uniform drag on all motions.  We end, in \S 2.4,  with a brief discussion
of the implications of this picture for observations of turbulence in the ISM.

\subsection{Neutral-ion damping}

The role of neutral-ion damping in MHD turbulence has been discussed
previously in the context of the ISM \citep[in particular, see][]{S91,MS97}.
The basic conclusion was that neutral fluid heating is a plausible sink
for the turbulent energy revealed through measurements of interstellar
scintillation.  Here we are concerned instead with how a neutral
gas component will modify the turbulent power spectrum.  The most
relevant observational point is that the ISM turbulent power
spectrum has no strong features at wavelengths where neutral-ion
coupling would be expected to play a dominant role \citep{ARS95}.
Instead, the power spectrum extends to very small scales ($<10^8$ cm)
in an approximate power law.
Qualitatively, this suggests that stochastic reconnection can
take place even in partially neutral plasmas.  However, several
basic questions remain unanswered.
Previous work on turbulence in the ISM has not included
a discussion of the most plausible model for MHD turbulence (although
a simple hydrodynamic model was addressed, which is remarkably
close to the model we use here).  Also, we need to understand
why neutral damping fails to produce a strong signature in the
ISM, or at least in the diffuse ionized component of the ISM, before
we can construct a general model for its role in partially ionized
plasmas.

\subsubsection{The Goldreich-Sridhar model}

The GS95 model of strong MHD turbulence is based on the notion of a Kolmogorov-like
cascade with an anisotropy imposed by the large scale magnetic field.
The exact degree of anisotropy follows from an average balance between
hydrodynamic and magnetic forces.
Eddies on a given scale are characterized by a wavenumber perpendicular
to the mean magnetic field direction, $k_{\perp}$, and a parallel
wavenumber, $k_{\|}$, such that the rate of eddy turnover time is
equal to the rate wave propagation along magnetic field, i.e.
\be
k_{\perp}v_k\approx k_{\|}V_A~~~,
\label{balance}
\ee
where $v_k$ is the typical velocity at the scale characterized
by the wavenumber $(k_{\|},k_{\perp})$.  As $v_k$ goes down with the
increase of $k_{\bot}$ this condition implies
eddies that are elongated along the field direction, and become
more elongated as we go to smaller scales.  In order to simplify our
notation, we will refer to $k_{\perp}$ below as $k$.

If energy is injected isotropically on some scale $l$, with $v_l\le V_A$, then
the cascade will begin in a regime of weak turbulence, in which motions can
be characterized as weakly interacting waves, with a frequency $\omega=k_{\|}V_A\sim $constant and
a nonlinear decay rate (see discussion in a review by Cho, Lazarian \& 
Vishniac 2003)
\be
\tau_{nl}^{-1} \sim {k^2v_k^2l\over V_A},
\ee
so that conservation of energy implies
\be
v_k\sim v_l(kl)^{-1/2}.
\ee
For $\tau_{nl}^{-1}$ less than the wave frequency 
the corresponding scale-dependent diffusion coefficient will be (LV99) 
\be
D_k\sim {v_k^2\over\omega^2\tau_{nl}}\sim {v_l^4l\over V_A^3},
\ee
which is actually independent of scale.  The decrease in wave motions at 
larger $k$ is
balanced by the decrease in the coherence time of the waves. 
Here we will ignore the contribution of these scales to turbulent 
diffusivity in
favor of the contribution from smaller, strongly turbulent, scales.

The weak turbulent cascade ends at a scale $k_T$ where equation (\ref{balance})
is satisfied.  At this scale
\be
v_T\sim {v_l^2\over V_A}, \hbox{\ and\ } k_Tl\sim \left({V_A\over v_l}\right)^2.
\ee
At larger $k$ the strong turbulence model applies and (see GS95)
\be
k_{\|} \approx l^{-1} \left({k\over k_T}\right)^{2/3}.
\label{k_p2m}
\ee
The rate of turbulent energy transfer is $k_{\|}V_A$, which means
\be
\tau_{nl}^{-1}\approx {V_A\over l}\left({k\over k_T}\right)^{2/3},
\label{tau2m}
\ee
while the rms turbulent fluid velocity is given by
\be
v_k\approx v_T \left({k\over k_T}\right)^{-1/3}.
\label{velm}
\ee
The magnetic field perturbations, $b_k$, are
\be
b_k\approx v_k (4\pi\rho)^{1/2}.
\label{belm}
\ee

This model presupposes that the turbulent velocities are subalfvenic,
and we adopt this assumption in the rest of this paper.  This is
less restrictive than it might appear, since as long as there is
some scale $l'$ in the turbulent cascade where $v_{l'}\sim V_A$ we
can take $l=l'\approx k_T^{-1}$, $v_T=v_l=V_A$ and use this model of turbulence
for all smaller scales. 
Moreover, if the turbulent energy is larger than the magnetic
energy, we can expect rapid growth of the magnetic field through the
turbulent dynamo, that is, the generation of a disordered field in
rough equipartition with the turbulent kinetic energy (see, however,
Schekochihin et al. 2003). 

The Goldreich-Sridhar scalings can be easily understood. They reflect
the fact that on small scales it is difficult to bend magnetic field lines,
but it is still easy to mix them up.

In a fully ionized astrophysical plasma, shear viscosity
is generally less important than resistivity
in damping MHD turbulence.  In a partially neutral medium a combination
of neutral particle viscosity and ion-neutral collisional coupling
drives damping.  Following
GS95 we concentrate on the diffusion of momentum across field lines.

\subsubsection{Ion-neutral decoupling: theoretical considerations}

The preceding discussion assumes that the plasma is a single, tightly coupled,
fluid with negligible viscosity.  Many astrophysical fluids are partially ionized, the
neutrals are imperfectly coupled to the ions. An obvious consequence is a substantially
increased viscosity since the neutrals can cross magnetic field lines.

The coupling between ions and neutrals is determined by
the rate of ion-neutral collisions, which is
\be
t_{in}^{-1}={m_n\over m_n+m_i} n_n\langle v_r \sigma_{in}\rangle,
\ee
where $v_r$ is the ion-neutral relative velocity, $\sigma_{in}$ is
the ion-neutral collisional cross section, $m_i$ and $m_n$ are the
typical ion and neutral masses, $n_n$ is the neutral number density,
and angular brackets denote averaging.  From \citep{DRD83}
we adopt ${\langle v_r \sigma_{in}\rangle}\approx 1.9\times 10^{-9}$
cm$^3$ s$^{-1}$. The rate at which neutrals exchange momentum with
ions is $t_{ni}^{-1}= t_{in}^{-1}\rho_i/\rho_n$.

If the mean free path for a neutral particle, $l_n$,
in a partially ionized gas
with density $n_{tot}=n_n+n_i$ is much less than
the size of the eddies under
consideration, i.e. $l_n k\ll 1$, the damping time is
\be
t_{damp}\sim
\nu_n^{-1} k^{-2}\sim \left(\frac{n_{tot}}{n_n}\right)(l_n c_n)^{-1} k^{-2}~~~,
\label{tdamp}
\ee
where $\nu_n$ is the effective viscosity produced by neutrals and
$c_n$ is the sound speed in the neutrals
\footnote{The viscosity 
across magnetic field lines, due to ion-ion collisions,
is typically small
as ion motions are constrained by the magnetic field.  For
a collision rate much smaller than the ion cyclotron frequency
the ratio of ion perpendicular viscosity to resistivity is a few times
the ratio of the ion thermal pressure to the magnetic pressure.
In the collisional limit this is multiplied by a factor
of $(\Omega_i/\nu_i)^2\ll 1$, and resistivity efficiently
dissipates magnetic field perturbations on scales greater
than the viscous damping scale.
The drag coefficient for neutral-neutral collisions is
$\sim 1.5\times10^{-10} T^{1/3}$ cm$^3$ s$^{-1}$ 
with $T$ measured in Kelvins
\citep{S78},
so collisions with other neutrals will dominate
for $n_i/n_n$ less than $\sim 0.08 T^{1/3}$.}.

Consider first a mostly neutral gas (i.e. $n_i\gg n_n$). 
Turbulence can cascade to small scales if the
turbulent eddy rate $\tau^{-1}\sim k v_k$ is larger than 
the viscous damping rate $t_{damping}^{-1}$. In a partially ionized
gas the one fluid approximation is valid if neutrals experience multiple collisions
with ions in an eddy turnover time, i.e. $t_{ni}^{-1}>\tau^{-1}$.
Therefore, MHD turbulence will exhibit Goldreich-Sridhar scaling
up to the damping scale if  $t_{ni}^{-1}>t_{damping}^{-1}$. On
the other hand, if
$t_{ni}^{-1}<t_{damping}^{-1}$ neutrals will not follow the ions and
magnetic fields. Instead they will form a {\it hydrodynamic} cascade
as soon as the neutral-ion collisional rate
$t_{ni}^{-1}$ is of the order of the eddy turnover time
$\tau$. At this scale, and all smaller scales,  the ionic fluid motions 
will damp at
the rate of ion-neutral collisions $t_{in}^{-1}$, which
in mostly neutral gas is much larger than $t_{ni}^{-1}$.

What will happen to the magnetic fields when ionic motions are damped?
Collisions between the ions and the neutral particles will prevent magnetic
tension from straightening the field lines efficiently.
Instead they will be moved,
entangled and stretched by large undamped eddies. As the eddy turnover
rate increases with the decrease of the scale, the marginally damped
eddies at the damping scale will the most important.
This implies a picture very different from the
Goldreich-Sridhar cascade.

If neutrals constitute a tiny impurity in the ionized plasma,
it is clear that they cannot affect the MHD cascade. We shal
quantify this intuitive picture below.  
  
\subsubsection{Ion-neutral decoupling: a simple model}

Consider first a toy model of ion-neutral interaction.
 If we ignore viscous damping, the equations for the
ions and neutrals are:
\be
{v_i\over\tau} ={v_n-v_i\over t_{in}} -{\rho\over\rho_i}
\omega_A^2\tau v_i+F_i~~~,
\label{memovi}
\ee
and
\be
{v_n\over\tau} ={\rho_i\over\rho_n}{v_i-v_n\over t_{in}}+F_n~~~.
\label{memovn}
\ee
For simplicity we have used $F_i$ and $F_n$ to denote both pressure 
forces and the nonlinear turbulent accelerations for the ions and neutrals, 
respectively.  

To linear order equations, and ignoring sound waves, (\ref{memovi}) and (\ref{memovn}) give
the dispersion relation for Alfv\'en waves in a partially
ionized plasma \citep{M77}:
\be
{t_{in}\over\tau}\left(1+(\omega_A\tau)^2{\rho\over\rho_i}\right)
={-1\over 1+\aleph}~~~,
\label{ldis}
\ee
where $\aleph$ is a dynamical coupling parameter, defined by
\be
\aleph\equiv {\rho_i\over\rho_n}{\tau\over t_{in}}~~~.
\ee
When the coupling is very tight, $\aleph\gg 1$, and we have the usual relation
for Alfv\'en and pseudo-Alfv\'en modes in a single fluid,
\be
\tau^{-2}=-\omega_A^2~~~,
\label{alf}
\ee
where $\tau$ is imaginary.

As the turbulence cascades to smaller scales, $\tau$, and
consequently, $\aleph$, decreases.
If we express the value of $\aleph$ at the viscous damping scale
as $\aleph_c$, then we have two obvious alternatives.
Either $\aleph_c>1$ and two-fluid effects are negligible right up to
the damping scale, or $\aleph_c<1$ and the ions and neutrals decouple
in the middle of the turbulent cascade.  In order to calculate
$\aleph_c$ we need to calculate viscous damping in the large $\aleph$
limit.

Combining Eqs~(\ref{k_p2m})
and
(\ref{tdamp}) we get
\be
{t_{damp}\over\tau}\sim f_n^{-1} \left(\frac{v_l}{V_A}\right)^{1/3}
\left(\frac{l_n}{l}\right)^{1/3} \left(\frac{v_l}{c_n}\right)
(l_n k)^{-4/3}~~~,
\label{eq.4}
\ee
where $c_n$ is the sound speed and $f_n$ is the neutral fraction.
For most of our applications we will have $f_n\sim 1$.
The damping scale, $k_c^{-1}$, is defined by $t_{damp}\sim \tau_k$, so that
\be
k_c\sim l_n^{-1}\left(\frac{v_l}{c_n}\right)^{3/4}
\left(\frac{v_l}{V_A}\right)^{1/4}\left(\frac{l_n}{l}\right)^{1/4}
f_n^{-3/4},
\label{4.2.1}
\ee
\be
\tau_c^{-1}\sim k_c^2f_nc_nl_n\sim
\left(\frac{c_n}{l_n}\right)\left(\frac{v_l}{c_n}\right)^{3/2}
\left(\frac{l_n}{l}\right)^{1/2} \left(\frac{v_l}{V_A}\right)^{1/2}
f_n^{-1/2},
\label{4.2.2}
\ee
\be
k_{\|,c}\sim \tau^{-1}_s V_A^{-1}\sim l_n^{-1}
\left(\frac{v_l}{c_n}\right)^{1/2}\left(\frac{l_n}{l}\right)^{1/2}
\left(\frac{v_l}{V_A}\right)^{3/2}f_n^{-1/2},
\label{4.2.3}
\ee
and
\be
v_c\sim v_l
\left(\frac{l_n}{l}\right)^{1/4}
\left(\frac{c_n}{V_A}\right)^{1/4}f_n^{1/4},
\label{4.2.4}
\ee

{}From the definition of 
$\aleph_c\equiv \aleph (\tau_c)$
we see
that
\be
\aleph_c\sim f_n^{1/2}
\left({l\over l_n}\right)^{1/2} \left({\rho_i l_n\over \rho_n t_{in}c_n}\right)
\left({c_n\over v_l}\right)^{3/2}\left({V_A\over v_l}\right)^{1/2}.
\label{als}
\ee

Equation (\ref{als}) seems to imply that $\aleph_c$ must always be greater than
one, but a closer examination suggests that for $\rho_i\ll \rho_n$ the third
term on the right hand side can be small enough to offset the second
term.  In fact, the value of $\aleph_c$ has to be determined for each
situation.  When $\aleph_c\ll 1$  equations (\ref{4.2.1}) through (\ref{4.2.4})
have no physical meaning, since
the plasma stops behaving as a single fluid when $\aleph$ drops below
one.  Nevertheless, we can still use these expressions as useful parameterizations
of the turbulent cascade.  In particular, if $\aleph_c<1$ then the scale of
decoupling, when $\aleph=1$, is given by
\be
k'\sim k_c\aleph_c^{3/2},
\label{decoup1}
\ee
while at higher wavenumbers equation (\ref{ldis}) becomes
\be
{t_{in}\over\tau}\left(1+\omega_A^2\tau^2{\rho\over\rho_i}\right)=-1~~~.
\label{memo6}
\ee
This expression has two roots. When the collision rate is so
small that the ions
and neutrals are completely decoupled we have usual
dispersion relation for Alfv\'en waves
\be
\tau^{-2}\sim -\omega_A^2 \rho/\rho_i~~~.
\label{decouple1}
\ee
When the collision rate is large the two roots are
\be
\tau^{-1}\sim -\omega_A^2t_{in}{\rho\over\rho_i},~~~ -{1\over t_{in}}.
\label{memo7}
\ee
The latter root corresponds to the case where magnetic forces are
negligible and the former is the usual ambipolar diffusion rate, when the magnetic
field pushes the ions through a neutral background.  Neither limit
is appropriate for hydrodynamic turbulence in the neutral fluid,
in which case $v_n$, $\tau$, and ${\bf k}$ are imposed by
the turbulent cascade.  From equation (\ref{memovn}) we see that
if the ions are prevented from moving by the magnetic field then
we get a damping rate $\sim ({\rho_i/\rho_n})t_{in}^{-1}$, which
will be negligible once the hydrodynamic cascade has proceeded
to eddies with turn over rates larger than the decoupling rate.
Since the hydrodynamic eddies will be approximately isotropic, equation
(\ref{memovi}) guarantees that $v_i\ll v_n$ for all scales below
the decoupling scale.

The limit defined by equation (\ref{decouple1})
has been previously considered \citep[see, for example][]{KP69}.
We will finish this subsection by considering damping in this regime, and
the minimum neutral fraction required for neutral damping on any scale.

It is easy to see that equation (\ref{decouple1}) corresponds to the regime
$l_n k\gg 1$ and the waves are damped
at a rate of $t_{in}^{-1}$. Therefore,
\be
\tau_k^{-1}t_{damp}\sim
\left(\frac{v_l}{c_n}\right)
\left(\frac{l_n}{l}\right)^{1/3}
\left(\frac{v_l}{V_A}\right)^{1/3}
\left({c_nt_{in}\over l_n}\right)
(l_n k)^{2/3}.
\label{o1}
\ee
If we have a turbulent cascade in the ions, with $\tau_{nl}^{-1}t_{in}>1$, then
the cascade proceeds to very small scales, and neutral damping plays no role 
in its dynamics.
Although the ions are decoupled from the neutrals, we use the
usual value of $V_A$ in this expression since it appears only
in the combination $v_l^4/(lV_A)$, which is the energy cascade
per unit mass.
The ratio $c_nt_{in}/l_n$ is  roughly
\be
{c_nt_{in}\over l_n}={l_{in}\over l_{nn}}+
{c_it_{in}\over c_nt_{ni}}
\approx {l_{in}\over l_{nn}}+
{\rho_i\over\rho_n}\left(\frac{m_n}{m_i}\right)^{1/2}.
\ee
If the ions and neutral particles have the
same mass, this will be $\sim f_n^{-1}$ when the neutral
fraction is small, and a number of order unity otherwise.

Here we have assumed that the rms velocities of the ions and neutrals
are dominated by the thermal distribution.
Our treatment should still be
applicable when the turbulence is supersonic, but sub-Alfv\'enic 
(see
discussion of the point in Cho, Lazarian \& Vishniac 2002a),
as long as we replace $c_i$ and $c_n$ with the appropriate
rms turbulent velocities. This
is most likely to happen on large scales, so that $l_n k\ll 1$.
In this case the viscosity contributed by neutrals stays
the same so that equation (\ref{eq.4}) does not
require any modifications. In the opposite limit, when  $kl_n\gg 1$,
for supersonic turbulence $\langle v_i^2\rangle^{1/2}\approx V_A^{\star}$,
where  $V_A^{\star}$
is the Alfv\'en velocity when ions and neutrals are decoupled, i.e.
$V_A^{\star}=(n_n m_n +n_i m_i)^{1/2}/(n_i m_i)^{1/2} V_A$.
However an increase in the ion velocity will result in a proportional
increase in the drag coefficient, and a consequent drop in $t_{in}$.
Thus the condition provided by equation (\ref{o1}) does not change.

Equations (\ref{eq.4}) and (\ref{o1}) can be used to find the
minimal neutral fraction required to damp the turbulent cascade
at scales close to $l_n$.  Together they imply a maximum damping
rate for $kl_n\sim 1$. We see $\tau_k^{-1}t_{damp}\sim 1$
at that scale requires a small neutral fraction, so that
$l_n\sim v_nt_{ni}$.  Assuming a single thermal velocity we get
\be
f_n>f_{crit}=7\times 10^{-2}\left (\frac{l}{pc}\right)^{-1/3}
\left({v_T\over c_n}\right)^{2/3}\left({V_A\over c_n}\right)^{1/3}
n_{tot}^{-1/3}T^{1/6},
\label{fcrit}
\ee
as the minimal condition for dissipating an MHD turbulent cascade
through neutral particle viscosity.
Typically, turbulent motions
in the ISM are of order $c_n$ or greater, so only a small neutral
fraction is necessary to affect the turbulent cascade.
If there is a source of noise on scales
$\ll l_n$, such that damping given in equation (\ref{o1}) is ineffective,
then neutral friction can be ignored on all smaller scales.
However, we will show in the following pages that small scale turbulence
can appear in a partially neutral plasma even
in the absence of any small scale driving.

\subsection{The coupled regime: viscosity-damped turbulence}

In this subsection we consider strong MHD turbulence with strong
neutral particle damping, but on scales where the one-fluid
approximation remains valid.

In hydrodynamic turbulence viscosity sets a minimal scale for
motion, with an exponential suppression of motion on smaller
scales.  Below the viscous cutoff the kinetic energy contained in a
wavenumber band is
dissipated at that scale, instead of being transferred to smaller scales.
This means the end of the hydrodynamic cascade, but in MHD turbulence
this is not the end of magnetic structure 
evolution.
For viscosity much larger than resistivity,
$\nu\gg\eta$, there will be a broad range of
scales where viscosity is important but resistivity is not.  On these
scales magnetic field structures will be created through a
combination of large scale shear and the small scale motions generated
by magnetic tension.  As a result, we expect
a power-law tail in the energy distribution, rather than an exponential
cutoff.

Here we discuss a conservative model for the damped regime.
It is motivated by simulations by \citet[henceforth CLV02b]{CLV02b}
and is consistent with those simulations \citep[see also][]{CLV03b}.   
A complete
understanding of this regime will be deferred until
higher numerical resolution runs are available and a more detailed
theoretical treatment has been completed.  Our goal here is a model
which can serve as an approximate guide.  We will construct this
model using the notion of local interactions in phase space
\citep[cf.][]{Sp99}, a
constant cascade of energy to small scales, and a force balance
between magnetic
tension and viscous forces.  We add to this two ingredients
that are directly suggested by the simulations: a constant curvature
for the field lines and an intermittent magnetic field distribution.

To begin with we define a filling factor $\phi_k$, which is the
fraction of the volume containing strong magnetic field perturbations
with a scale $k^{-1}$.  We denote the velocity and perturbed
magnetic field inside these subvolumes with a ``$\hat{\ } $'' so
that
\be
v_k^2=\phi_k \hat v_k^2,
\ee
and
\be
b_k^2=\phi_k \hat b_k^2.
\ee

We note that at the critical damping scale, $k_c$ and
$k_{\|, c}$, the time-scale for inviscid turbulence is
\be
\tau^{-1}_s\approx k_c^2 \nu\approx k_{\|, c} V_A\approx k_c v_c
\approx k_c{b_c\over (4\pi\rho)^{1/2}}~~~.
\label{4.1}
\ee
Assuming we are in the coupled regime, we have $\aleph_c=\rho_i \tau_c/
(\rho_n t_{in})>1$.  Motions on this scale are marginally damped, so that
smaller scale structures will be continuously sheared at a rate $\tau_c^{-1}$.
These structures will reach a dynamic equilibrium if they generate a
comparable shear, that is
\be
k \hat v_k\sim \tau_c^{-1}.
\label{s1}
\ee
The magnetic energy will cascade to higher wavenumbers at this
same rate, so that
\be
b_k^2k\hat v_k\sim {b_c^2\over \tau_c}.
\label{s2}
\ee
Consequently, $b_k\sim b_c$ for $k>k_c$.

Next we assume that the curvature of the magnetic field lines
changes slowly, if at all, in the cascade.  This is consistent
with a picture in which the cascade is driven by repeated shearing
at the same scale.  It is also consistent with the numerical
work described in CLV02b, which yielded a constant $k_{\|}$
throughout the viscously damped nonlinear cascade.

Finally, we can balance viscous and magnetic tension forces
to find
\be
k^2\nu \hat v_k \sim k\nu\tau_c^{-1}\sim \max[\hat b_kk_c,B_0k_{\|,c}] \hat b_k
\sim k_c\hat b_k^2.
\label{s3}
\ee
This suggests a picture in which the cascade to larger wavenumbers consists
of an evolution to increasing gradients perpendicular to both the
mean field direction and the local perturbed field component.  This
implies rapidly increasing magnetic pressure gradients, but these
can compensated by plasma density fluctuations.

Combining our results we find
\be
\hat b_k \sim b_c \left({k\over k_c}\right)^{1/2}, b_k\sim b_c,
\label{4.5}
\ee
\be
\hat v_k\sim v_c \left({k_c\over k}\right), v_k\sim
v_c\left({k_c\over k}\right)^{3/2},
\label{4.6}
\ee
and
\be
\phi_k\sim {k_c\over k}.
\label{4.6a}
\ee

The associated velocities fall rapidly, although not exponentially.
In terms of one dimensional spectra we have
a magnetic energy spectrum $E^b(k)\sim k^{-1}$,
and a kinetic energy spectrum $E^v(k)\sim k^{-4}$. These scaling
laws should be compared
with $E^v\sim E^k\sim k^{-5/3}$ for the Goldreich-Sridhar predictions
for the inviscid regime.  The fact that the local magnetic perturbations 
increase,
albeit slowly, with increasing wavenumber implies that they will eventually
exceed the strength of the background field.  Further
research should clarify whether this is a problem.


If we compare this model to the simulations shown in CLV02b and the
analysis in \citet{CLV03b} we see that the magnetic field Fourier power spectrum 
in the simulations is slightly steeper than expected, i.e.
\be
E_B(k)\propto k^{-1.2\hbox{\ to\ }-1.3},
\ee
while the kinetic energy spectrum shows a stronger deviation
\be
E_v(k)\propto k^{\sim -4.5}.
\ee
The fact that the deviation in the exponent is twice as large for
$E_v$ as for $E_B$ is consistent with equation (\ref{s3}).  It remains
only to explain why the magnetic power spectrum is not actually flat.  The most
likely explanation is that since the power spectrum amplitude at a given $k$ 
is the Fourier transform of the correlation function, it represents an
integration of the correlation function over all scales $r$ less than 
$\sim k^{-1}$.
Since the energy is distributed logarithmically on all scales between 
the damping scale and the dissipation scale, this introduces a
factor $\sim\ln (k_{min}/k)$ to $E_B(k)$.  
The simulations have a range $\sim 20$ between the scale of viscous damping
and the scale of resistive damping,  so this bends the slope
of $E_B(k)$ downward by roughly the required amount. 

The functional form of $\phi_k$ is somewhat more difficult to
test.  In CLV03b we constructed filtered maps of $b$ by separating the
${\bf b}({\bf k})$ into broad bins. Plotting the cumulative
magnetic energy versus volume for each filtered map gives a
measure of the volume filling fraction as a function of scale.
We show the results in figure 2 (Fig.~4 in CLV03b).  Comparing
the concentration of magnetic energy as a function of scale
we see that, as expected, the concentration of magnetic power
increases at smaller scales.  Our results may have a slightly shallower 
dependence on wavenumber than predicted by equation (\ref{4.6a}),
but this depends on the value of the cumulative fractional magnetic 
energy we choose for comparison.

Finally, we note that the eddies in this regime are not described by
the linear solutions to equations (\ref{memovi}) and (\ref{memovn}).
The transfer of magnetic power from large to small scales
makes these solutions largely irrelevant.

\subsection{The decoupled, damped regime}

As we have seen, ion-neutral decoupling may happen either before
or after the turbulent
cascade reaches the viscous damping scale.  In both cases the cascade
will show similar behavior in the decoupled and damped regime, although the
details of the onset of this stage will differ.  We begin by discussing the
case where decoupling occurs below the viscous damping scale, that is, after
the cascade has entered the stage described in the preceding subsection.
We will discuss the case of early decoupling, $\aleph_c<1$, in the middle of
this subsection.  We will conclude this subsection by describing the termination
of the damped regime and the reappearance of a strong turbulent cascade involving
only charged particles.

When $\aleph_c>1$, decoupling occurs as a two step process.
First, the pressure support
from the neutral particles becomes ineffective and the energy cascade
is reduced by a factor of $(P_i/P_{tot})\sim (n_i/n_{tot})$.  At somewhat
smaller scales the difference between the ion and neutral velocities
becomes comparable to $v_k$ and the neutrals can be treated as
a static background.  We will rely on the same physical arguments used
successfully in the previous subsection to predict 
this part of the spectrum.  Currently numerical
simulations address only larger scales.

The first stage of decoupling occurs when the ambipolar diffusion rate
becomes comparable to $\tau_c^{-1}$ and the neutral particles infiltrate
the zones of intense magnetic field perturbations.
This happens at a wavenumber $k_p$ given by
\be
f_nk_p \hat b_p^2\sim {\rho_i\over t_{in}}{1\over k_p\tau_c},
\label{pdecoup}
\ee
where we use the subscript `p' to denote the pressure decoupling
scale.  Combining this criterion with equations  (\ref{4.1}) and (\ref{4.5})
we see that
\be
{k_p\over k_c} \sim \aleph_c^{1/3}.
\ee
When $f_n\sim 1$ the streaming of neutrals across the perturbed field lines
will be accompanied by significant dissipation, so that the
turbulent energy cascade is
reduced in amplitude by the factor by which the pressure support is
reduced, that is $\sim \rho_i/\rho$.   The volume filling factor
must also drop by this same factor in order to maintain the
condition expressed in equation (\ref{s3}).  
In our simplified model we will model the sharp drop in
$b_k$ and $\phi_k$ at $k_p$ as a discontinuity, although in
reality we expect a smooth transition.
It is important to note that there is no associated
discontinuity in the rms electron density fluctuations.  The volume
averaged strength of the magnetic pressure fluctuations drops by $\rho_i/\rho$
as the ions go from providing only a fraction ($\rho_i/\rho$) of the
compensating plasma pressure to supplying all of it.  The only
observational signal may be a change in higher order moments of the
scintillation statistics.

The neutrals and ions decouple entirely when the viscous drag coefficient
becomes comparable to the neutral drag term in the ion force equation.  This
sets in when
\be
\rho \nu k^2 \sim {\rho_i\over t_{in}},
\ee
or at
\be
k_d\approx k_c (f_n\aleph_c)^{1/2}.
\label{kd}
\ee
For $k>k_d$ the conservation of energy condition and force balance
equations become
\be
b_k^2 \sim b_c^2{\rho_i\over\rho},
\label{c1}
\ee
and
\be
{\rho_i\over t_{in}} {1\over k\tau_c} \sim k_c \hat b_k^2.
\label{c2}
\ee
This implies
\be
\hat b_k\sim (f_n\aleph_c)^{1/2}\left({k_c\over k}\right)^{1/2} b_c,
\label{bsp3}
\ee
\be
\phi_k\sim {k\over k_c} {\rho_i\over\rho}(f_n\aleph_c)^{-1}
\sim{k\over k_c}{t_{in}\over\tau_c},
\label{fsp3}
\ee
\be
v_k\sim v_c \left({k_c\over k}\right)^{1/2}\left({t_{in}\over\tau_c}\right)^{1/2}.
\label{vsp3}
\ee
Equations (\ref{4.5}), (\ref{kd}) and (\ref{bsp3})
imply a maximum local perturbed field strength, at wavenumbers $\sim k_d$, of
\be
b_{max}\sim b_c(f_n\aleph_c)^{1/4}.
\ee
Under most conditions $b_{max}$ will be only moderately larger than $b_c$.
We note that the volume filling fraction {\it increases} with wavenumber
in this regime. This prediction should be tested using a two fluid code.

How is picture modified when we consider $\aleph_c<1$?  The decoupling condition,
$\aleph\sim 1$, evaluated to obtain equation (\ref{decoup1}), is equivalent to
the pressure decoupling criterion given in equation (\ref{pdecoup}) in this case.
We see from equation (\ref{memo7}) that turbulent motions in the ions
are strongly damped below this limit, so that we are immediately in the
damped, pressure decoupled and dynamically decoupled limit.
Consequently, we expect an immediate drop in $\phi_k$ and $b_k$ at the decoupling
scale.  On smaller scales we can invoke force balance and energy conservation
to write:
\be
b_k^2 \sim b'^2{\rho_i\over\rho},
\label{d1}
\ee
and
\be
{\rho_i\over t_{in}} {1\over k\tau'} \sim k' \hat b_k^2,
\label{d2}
\ee
where $\tau'$ is the eddy turn over time at the scale $k'$ defined in
equation (\ref{decoup1}).
Equations (\ref{d1}) and (\ref{d2}) imply
\be
\hat b_k\sim b' \left(f_n{k'\over k}\right)^{1/2}\sim b_c\aleph_c^{1/4}
\left(f_n{k_c\over k}\right)^{1/2},
\label{bsp4}
\ee
\be
\phi_k\sim {\rho_i\over\rho_n}\left({k\over k'}\right)
\sim {\rho_i\over\rho_n}\left({k\over k_c}\right)\aleph_c^{-3/2},
\label{bsp4a}
\ee
and
\be
v_k\sim v' \left({k'\over k}\right)^{1/2}\left({\rho_i\over \rho_n}\right)^{1/2},
\label{vsp4a}
\ee
where
$v'$ and $b'$ are the rms velocities and magnetic field perturbation strengths
at the scale $k'$.  The differences between equations (\ref{bsp3})-(\ref{vsp3})
and equations (\ref{bsp4})-(\ref{vsp4a}) are due to the different stirring rates
imposed by the marginally damped eddies in the two different cases.

The persistence of the energy cascade to very small scales implies that
the magnetic field will revert to a strong turbulent
cascade once collisions with the neutral background are slower than the
magnetic dynamical evolution rate.  This sets in when
\be
{k \hat b_k\over (4\pi\rho_i)^{1/2}} \sim t_{in}^{-1}.
\label{onset}
\ee
{}From equations (\ref{bsp3}) and (\ref{bsp4}) we find that the onset of small
scale turbulence is at a wavenumber
\be
k_t\sim k_c \left({\rho_n\over\rho_i}\right)\aleph_c\min[1,\aleph_c^{1/2}].
\label{ktdef}
\ee
In both cases $k_t$ coincides with a filling fraction of order unity, so that
the distinction between $\hat b_k$ and $b_k$ disappears.
At this wavenumber small scale instabilities will cause $k_{\|}$ and $v_k$ to
jump sharply, while $b_k$ drops.  There will be a sharp mismatch between the
correlation time of the damped turbulence on slightly larger scales and the
eddy turnover time of the strong small scale turbulence, which will start at
$t_{in}$.  This suggests that the turbulent cascade will be driven
intermittently, with a buildup of small scale energy followed by
its rapid release.  The duty cycle for this process will be approximately
the ratio of $t_{in}$ to the eddy turn over time for the marginally damped
large scale or
\be
\epsilon\sim {\rho_i\over\rho_n} \min[1,\aleph_c^{-1}].
\label{inter}
\ee
During the active phase of the small scale turbulence the cascade rate will
be
\be
b(k_t)^2/t_{in}\sim \rho_n {v_l^4\over V_A l}\max[1,\aleph_c].
\ee
The corresponding turbulent spectrum will be
\be
v_k\sim v_l \left({v_l\over V_A}\right)^{1/3}(kl)^{-1/3}
\left({\rho_n\over\rho_i}\right)^{1/3}
\max[1,\aleph_c^{1/3}],
\label{vsp4}
\ee
and
\be
k_{\|}\sim {1\over l}(kl)^{2/3}\left({v_l\over V_A}\right)^{4/3}
\left({\rho_i\over\rho_n}\right)^{1/6}f_n^{1/3}
\max[1,\aleph_c^{1/3}],
\label{pp}
\ee
with $b_k\sim (4\pi\rho_i)^{1/2}v_k$.

These bursts of turbulence will terminate at some small
dissipative scale, due to either ohmic resistivity,
ion viscosity along the magnetic field lines or to
plasma effects in a collisionless medium.  The resistive
perpendicular wavenumber is set by the condition
$k_{res}\eta\sim v_k$, or
using equation (\ref{vsp4})
\be
k_{res}\sim l^{-1} \left({v_ll\over\eta}\right)^{3/4}
\left({v_l\rho_n\over V_A\rho_i}\right)^{1/4}
\max[1,\aleph_c^{1/4}].
\ee

For partially ionized gas in the interstellar medium
\citep[see][for a discussion of idealized interstellar phases]{DL98} the
collision rate
is much less than the ion cyclotron frequency and the ion Larmor radius
is greater than $k_{res}^{-1}$.  The condition for damping due to
parallel ion viscosity is
\be
k_{\|}^2 c_n l_i\ge k_{\|}V_A^*,
\ee
where, as before, $V_A^*$ is the Alfv\'en velocity for the ions
alone, and $k_{\|}l_i<1$.  In the limit where the plasma is
strongly magnetized and $\rho_i\ll\rho_n$ we expect $V_A^*>c_n$, so
that this criterion is never satisfied.  That is, damping due to
parallel transport is instead given by free-streaming along the field lines,
which has an associated damping rate which is always less than the
cascade time scale.  We have already noted that the perpendicular
ion viscosity
is, at most, only slightly greater than the resistivity. It follows that
the turbulent cascade is truncated at the Larmor radius, when
$kr_L\sim 1$.  In dense plasmas, like stars and
accretion disks, collisions dominate, ion viscosity is negligible,
and the cascade ends at the resistive scale.

\subsection{Turbulence in the ISM}

Interstellar medium is turbulent with turbulence spreading over a
range of scales from hundreds of parsec \citep[see][]{LP00,SL01}
to astronomical units (see Spangler 1999)
As \citet{ARS95} pointed out, the fact that interstellar scintillation
suggests a power law spectrum consistent with Kolmogorov turbulence
is already a strong indication that the observed scales are
connected by a power cascade.  The lack of any feature clearly
attributable to neutral particle damping can be seen as a counter-argument,
but the model sketched above implies that it will be difficult to
detect such a feature.  For example, although the average neutral fraction in
the diffuse ionized medium is uncertain, it probably lies somewhere
between a percent and a few times that.  From equation (\ref{fcrit})
we see that the critical neutral fraction
for this gas is on the order of several percent, depending on the
local Mach number, so
$f_n\aleph_c$ is at most of order unity.  For low Mach numbers
one might expect this to translate
into an exponential suppression by a factor of order unity,
since the eddy turnover rate will drop with the amplitude
of motions on a given scale, allowing for more effective damping.
However, since the electron density fluctuations will trace the magnetic
pressure
fluctuations caused by pseudo-Alfv\'en modes within the turbulent
cascade, the mean square variance of electron density on small
scales, $k_{large}>k_t$, will look like an extrapolation of the same
quantity from large scales, $k_{small}<\min[k',k_c]$,
but reduced by a factor of
\be
\epsilon \left({\rho\over\rho_i}\right){b(k_{large})^2\over b(k_{small})^2}
\left({k_{large}\over k_{small}}\right)^{2/3}
\sim \left({\rho_i\over\rho_n}\right)^{1/3}\min[1,\aleph_c^{-1/3}].
\ee
This will be a very modest reduction.  The scintillation power
spectrum will look like a fairly continuous power law from large
to small scales, but with a range of intermediate scales where
it goes flat (from $k_c$ or $k'$ to $k_t$) before dropping
sharply to an extrapolation of its large scale behavior.
The dynamic range of this flat region will be
$\sim (\rho_n/\rho_i)\max[1,\aleph_c]$.  Again, in the diffuse
ionized medium this will correspond to a factor of order unity
(or less) in length scales, which will not produce an observational
signature.  The warm neutral medium should show a more
pronounced shoulder in the power spectrum, although given
the heterogeneity of the local interstellar
medium, detecting the signal of neutral damping using electron density fluctuations
still represents a challenge. The coldest, and densest, phases of the ISM
have a negligible impact on the scintillation measurements.

In Table~1 we show critical scales and parameters for some idealized phases 
of the ISM, taken from \citet{DL98}.  The most obvious point is that the
different regimes discussed in this section cover a very modest range
of scales.  Ignoring the parallel scales, which may not leave a clear
observational signal, the entire range of damped scales covers only two
orders of magnitude in the cold and warm neutral phases of the ISM.
For the bulk of the ISM, the turbulent cascade
consists largely of a large scale cascade involving all the particles, and
a smaller scale cascade involving only the ions.  (We have not considered the
role of charged grains in this analysis.)  Current simulations of strongly
damped MHD turbulence, which apply only to the range between $k_c$ and $k_p$,
already have {\it more} dynamic range than in the ISM itself.  The small
wavelength limit on ISM turbulence, set by the Larmor radius, is only
weakly dependent on local conditions.

In magnetically dominated environments, e.g. in molecular clouds, compressibility 
can be important. This will extend the range of scales over which the turbulence
will lie in the viscosity damped regime. 

\section{Reconnection Rates}

We are now in a position to rederive the speed of reconnection
in turbulent magnetized plasmas, including the effects of a neutral
component.  Below we briefly consider highly ionized plasmas, extending
our previous analysis and highlighting steps in our derivation that
will be affected by neutral particles.  In \S 3.2 we consider stochastic
reconnection in turbulent plasmas with a small ionized fraction.

\subsection{Highly ionized plasmas}

For $f_n<f_{crit}$, as defined in equation (\ref{fcrit}),
the ion-neutral coupling is dynamically insignificant on all scales.
At sufficiently small scales a modest fraction of the energy cascade will
go into heating the neutral gas, but given the approximate nature of
our discussion we will ignore the resulting corrections to
$v_k$.  This limit is equivalent to the fully ionized case,
described in LV99.  Here we will briefly review the major results
from that paper, since we need to generalize them in order to
calculate reconnection speeds for partially neutral plasmas.

The basic geometric constraint on reconnection speeds is
\be
V_{rec}={\Delta(L)\over L} V_{eject},
\label{v1}
\ee
where $\Delta(L)$ is the width of the ejection surface for a current
sheet of length $L$.  In the absence of viscosity or neutral friction
the ejection speed is $\sim V_A$. The Sweet-Parker rate comes from
taking $\Delta\sim \eta/V_{rec}$.  In a collisionless plasma
$\Delta$ can be $\sim r_L$, the ion Larmor radius, when this is larger
than the resistive layer.  Both of these
estimates ignore the role that current sheet instabilities may play in
increasing $\Delta(L)$.  Here we are principally concerned with another
effect, which is that as the ejected plasma moves out of the reconnection
region, along magnetic field lines, the size of $\Delta(L)$ will
increase as the magnetic fields lines diffuse away from the current sheet.

In the presence of noise individual magnetic field lines will stay within
the current sheet for some distance, $L_{min}\ll L$, whose value will
depend on the current sheet thickness and the amplitude of the noise.
These individual patches of flux will reconnect at some speed
$V_{rec,local}$ obtained by substituting $L_{min}$ in place of
$L$ in equation (\ref{v1}).  In the presence of noise, the current sheet
contains many independent flux elements, all reconnecting simultaneously,
so that we have an upper limit on the reconnection speed given by
$\sim (L/L_{min}) V_{rec,local}$.  Since the geometric constraint
on the reconnection speed can be applied on any scale between
$L$ and $L_{min}$ this implies
\be
V_{rec}< {\Delta(L')L\over (L')^2} V_{eject}(L'),\hskip 1cm \hbox{for all\ }
L'\in[L_{min},L],
\label{v2}
\ee
where $\Delta(L')$ is the distance field lines wander perpendicular to the
large scale magnetic field direction within a distance $L'$ along the
field lines.  For a turbulent cascade the diffusion length $\Delta$ is
the solution to
\be
{d \Delta^2\over dx} = D_B(\Delta)
\label{v3}
\ee
where $D_B$ defines the stochastic diffusion of field lines separated
by a distance $\Delta$.  Assuming that the field line stochasticity is
caused by turbulence in the medium, the functional form of $D_B$ depends on the
nature of the turbulence, and is likely to be a function of scale.  
For randomly diffusing field lines this is
\be
D_B(\Delta)\sim \max[\left({b_k\over B_0}\right)^2k_{\|}^{-1}]
\hskip 1cm \hbox{for all\ }k\Delta\ge 1.
\label{diff1}
\ee
where $k_{\|}(k)$ is the parallel wavenumber as a function of the
perpendicular wavenumber.  In the presence of strong turbulence this
expression is $k^{-2}k_{\|}(k)\propto k^{-4/3}$.
For long wavelength perturbations, $k\Delta\le1$,
there can be a contribution which is reduced from its value at the scale
$k^{-1}$ by factor $(k\Delta)^2$.  Assuming strong turbulence and taking
$\Delta k\sim 1$ we see from equations (\ref{v3}) and (\ref{diff1}) that
we expect $\Delta\propto k_{\|}^{-3/2}\sim L'^{3/2}$.  In other words, the
rms separation of field lines should grow as the distance along the direction
of the mean field to the $3/2$ power.  In figure 3 we plot rms separations
for a variety of initial separations in a $256^3$ cubed hyperviscosity
and hyperresistivity simulation.  The details of the simulation can be
found in CLV02a.  We see that the separation grows as distance along
the field lines to the $3/2$ power, as expected, subject
only to a time offset caused by the finite initial separation.
Also as expected, above the dominant eddy scales the curves roll over into the
usual $L^{1/2}$ law.  Independent, but similar, simulations are 
described in \citet{SMCM02} who found similar results.

In LV99, and here, we are concerned with the effect of magnetic field
line diffusion on reconnection.  However, this same effect removes
the suppression of thermal conduction perpendicular to large scale field 
lines in diffuse magnetized plasmas \citep{MN01,CLHKKM03, MCB03, CM03}.
This has had a dramatic effect on our understanding of the thermal
structure and history of hot plasmas in galaxy clusters, which has not 
yet reached a final resolution.

For an ionized plasma with negligible viscosity, $V_{eject}\sim V_A$ at
all scales.  Combining equations (\ref{k_p2m}), (\ref{velm}), (\ref{belm}),
(\ref{v2}) and (\ref{diff1})
we find that the most restrictive limit on $V_{rec}$ comes from $L'\sim L$
and gives
\be
V_{rec}< v_T\min\left[\left({L\over l}\right)^{1/2},
\left({l\over L}\right)^{1/2}\right].
\label{eq:lim2a}
\ee
In a real ionized plasma the viscosity is {\it not} necessarily negligible.
When the particle collision rate is low the 
 perpendicular viscosity
 will be larger than the resistivity
by a factor of several times the ratio of gas pressure to magnetic
field pressure.  Under relevant astrophysical circumstances this will
be factor of order $10$, raising the possibility that viscosity may
reduce magnetic field line diffusion at scales slightly larger than
the scale of resistive dissipation
However, since the strongest
constraint comes from the largest scales, this will not affect 
the constraint given in equation (\ref{eq:lim2a}).

The claim that this is an actual estimate of the reconnection speed,
rather than an upper limit, follows from the absence of any other
important constraints on reconnection.  In this case the most obvious
alternative constraint is
that individual flux elements must reconnect many times
after their initial reconnection.  If subsequent reconnection events
are slow, then the current sheet will quickly evolve into a tangled
mass of reconnected magnetic field lines.  If the scale of the current
sheet is set by the ion Larmor radius, then this will define the
scale for all secondary reconnection events, and the small scale
reconnection speed will be $\sim V_A$.  The case where resistivity
defines the width of the current sheet is more complicated.  In LV99
we showed that we can consider secondary reconnection events to be
similar to the large scale reconnection, then used a self-similarity
argument to show that secondary reconnection was fast.  In either
case we find that equation (\ref{eq:lim2a}) is, as advertised, the
actual reconnection speed for field lines in turbulent background.

\subsection{Partially ionized plasmas}

As we have shown in \S 2, a turbulent cascade in a partially ionized plasma
is considerably more complicated than its counterpart in the fully ionized
case.  In particular,
the limit on the reconnection speed expressed in equation (\ref{v2}) no
longer increases monotonically with wavenumber, and equation (\ref{eq:lim2a})
may not be a fair estimate of the reconnection speed.  We need to consider
the full range of dynamical regimes, and search for minima in the reconnection
speed limit given by equation (\ref{v2}).  This requires estimates for
$V_{eject}$ and $D_B(\Delta)$ over the full range of the turbulent cascade.
We have already seen that for $k<k_c\min[1,\aleph_c^{3/2}]$ the most stringent
limit comes from $L'\sim L$, which gives a fairly generous upper limit
on the reconnection speed.

We begin by noting that when viscous drag is important, the ejection speed
from a volume of thickness $\Delta$ and length $L'$ is
\be
V_{eject}\approx {V_A^2\tau_c\over L'}\max[(k_c\Delta)^2, (f_n\aleph_c)^{-1}].
\label{eject}
\ee
At large scales, $k<k_c$, the plasma acts as a single fluid and we
have $V_{eject}\approx V_A$.  At very small scales, when the ions are
completely uncoupled from the neutrals, we get
$V_{eject}\approx V_A(\rho/\rho_i)^{1/2}$.

For $k>k_t$ a rescaled turbulent cascade
emerges.  As long as the current sheet thickness is smaller than $k^{-1}$,
a rescaled version of our previous analysis emerges.  In particular,
we expect the most important constraint on the reconnection speed to
arise from the smallest wavenumbers in this regime, $k\sim k_t$.  Due
to the intermittent nature of the turbulence we have a factor $\epsilon$
multiplying the usual expression.  Using equations
(\ref{ktdef}) and (\ref{inter}) we get
\be
D_B(k>k_t)\sim \epsilon k_t^{-2}k_{\|,t}\left({k_t\over k}\right)^{4/3}\approx
\left({\rho_i\over\rho_n}\right)^{1/2}\left({t_{in}\over\tau_c}\right)^2
{k_{\|,c}\over k_c^2}\left({k_t\over k}\right)^{4/3},
\label{dbd1}
\ee
which is always much less than the diffusion coefficient associated with
the viscous damping scale.  However, since it applies to much smaller
scales it still has physical significance.

The diffusion coefficient for the damped scales is less obvious.
The first complication is that since we are interested in the width
of the ejection zone, the use of an rms value for $\Delta$ is not
obviously correct.  The median value would be more appropriate.
Since we are not concerned with constants of order unity here
the distinction is unimportant if the distribution function for
field line separations is a smooth function with a single peak.
Although it is not obvious that this condition is satisfied in
the highly intermittent viscously damped regime, we see from
figure 4 that the median and rms values of field lines separation
seem to track one another as a function of distance along
the field lines.  The median is a factor of few lower.

The second complication is that our model of the viscously damped
regime does not predict the value of $D_B(k)$.  Since the curvature
wavenumber of the field lines is constant as a function of scale, we
can estimate the diffusion coefficient as
\be
D_B(k)\sim \left({k_{\|,c}\over k_c}\right)^2l_c \phi_k
\label{dbv}
\ee
where $l_c$ is the distance a field line remains within an intermittent
structure and $\phi_k$ is, as before, the filling factor of such structures.
The value of $l_c$ is uncertain, but goes to $k_{\|,c}^{-1}$ for $k=k_c$,
and must fall much more steeply than $k^{-1/3}$ to be consistent with
the sharply reduced value of $D_B(k)$, relative to the undamped case, seen
in figure 3.  Based on a simple geometric picture, we will assume
\be
l_c\approx k_{\|,c}^{-1}{k_c\over k}.
\label{lc}
\ee
This implies that
\be
D_B(k)\sim {k_{\|,c}\over k_c^2}\left({k_c\over k}\right)^2,
\ee
in the damped, viscously coupled regime with $\aleph_c>1$.
However, this is also the level
of diffusion we would expect from the shear imposed at $k=k_c$. Consequently,
the sharp drop in $\phi_k$ at the scale of pressure decoupling does not
affect field line diffusion.  In this regime the value of $\Delta$
rises exponentially, with a constant $L'\sim k_{\|,c}^{-1}$.  We see in
figure 4 that the simulations are consistent with comparable contributions
to the diffusion coming from large and small scales, and with an
exponential rise in $\Delta$.  The sharpest
constraint on $V_{rec}$ will come from the minimum $\Delta$ in this regime.

Our claim that $l_ck_{\|,c}\ll 1$  is based on the fact that the growth
in field line separations is not consistent with the idea that a small
fraction of field lines diverge sharply from their neighbors for a
distance $\sim k_{\|,c}^{-1}$.  We can get some idea of the true situation
by considering the full distribution of field line separations, given in
figure 5.  We see that while the separations certainly do not show a
gaussian distribution, they do show that field lines join the growing tail of
large separations at a distance which is typically shorter than $k_{\|,c}^{-1}$.

Equations (\ref{dbv}) and (\ref{lc}) imply that for $k>k_d$ the
diffusion coefficient will go to a constant value, which persists
until $k=k_t$ and $\phi_k=1$.  In this regime, $k_d<k<k_t$, we can
use equation (\ref{ktdef}) and find
\be
D_B(k)\approx {k_{\|,c}\over \min[k_t,k']k_c}\approx
{k_{\|,c}\over k_c^2}{t_{in}\over\tau_c}\max[1,\aleph_c^{-1}],
\label{dbd}
\ee
However,
this value is only relevant when, for some $k>k_d$,
it is greater than the contribution from the
large scale shear given by $\sim k_{\|,c}\Delta^2$ (for $\aleph_c>1$)
or $\sim k_{\|}'\Delta^2$ (for $\aleph_c<1$).
Once this value of $D_B$ applies we get $\Delta\propto L'^{1/2}$ and
the limit on $V_{rec}$ will increase with $k$.  The minimum upper
limit on $V_{rec}$ is at a scale
\be
\Delta \approx k_c^{-1} \left({t_{in}\over\tau_c}\right)^{1/2}
\max[1,\aleph_c^{-1/2}],
\label{dmin1}
\ee
with a corresponding $L'$ of
\be
L'\approx k_{\|,c}^{-1}\max[1,\aleph_c^{-1}].
\label{dmin1a}
\ee
We see from equation (\ref{eject}) that the corresponding ejection
velocity is
\be
V_{eject}\approx {V_A\over f_n}\min[1,\aleph_c^{-1}]
\label{eject1}
\ee

Comparing equations (\ref{dbd1}) and (\ref{dbd}) we that there is a
drop in $D_B(k)$ at $k=k_t$, implying that on some smaller scale
where the larger scale shearing is comparable to the turbulent
diffusion we have another local minimum in the upper limit for $V_{rec}$.
This happens at a field line separation of
\be
\Delta\approx k_t^{-1}\left({t_{in}\over\tau_c}\right)^{3/2}
\left({\rho_i\over\rho}\right)^{3/4}\min[1,\aleph_c^{3/2}]
\approx k_c^{-1}\left({t_{in}\over\tau_c}\right)^{5/2}
\left({\rho_i\over\rho_n}\right)^{3/4}\min[1,\aleph_c],
\label{dmin2}
\ee
with a corresponding $L'$ of
\be
L'\approx k_{\|,c}^{-1} \left({t_{in}\over\tau_c}\right).
\label{dmin2a}
\ee
On such small scales the ejection speed will be $V_A(\rho/\rho_i)^{1/2}$.

We are now in a position to evaluate the upper limits on the reconnection
speed.  We begin with the one set at very small scales.  Using equations
(\ref{k_p2m}), (\ref{tau2m}), (\ref{v2}), (\ref{dmin2}),
and (\ref{dmin2a}) we find that
\be
V_{rec}<v_l{L\over l}
\left({\rho_i\over\rho}\right)^{1/4}f_n^{-3/4}
\left({l\over v_lt_{in}}\right)^{1/2} \min[1,\aleph_c^{-1}].
\label{llow}
\ee
This is generally less restrictive than the limit given in equation (\ref{eq:lim2a}).

The limit on the reconnection speed from intermediate scales can be
found from equations (\ref{4.2.1}), (\ref{4.2.3}), (\ref{als}),
(\ref{v2}), (\ref{dmin1}),
(\ref{dmin1a}), and (\ref{eject1}).  In this case we find
\be
V_{rec}<v_T{L\over l}
{\rho_i\over\rho}f_n^{-2}
\left({l\over V_At_{in}}\right)^{1/2} \min[\aleph_c^{-2},\aleph_c^{1/2}].
\label{lim2b}
\ee
This is potentially more restrictive than equation (\ref{eq:lim2a}), and
certainly more restrictive than equation (\ref{llow}), and needs
to be evaluated for specific circumstances.  This limit applies whenever
the current sheet is narrower than the value of $\Delta$ given
in equation (\ref{dmin1}).  For the specific examples we consider
in the next section this is usually the case.

\section{Reconnection in Various Phases of ISM}

All common phases of the ISM are strongly collisionless, so that
ambipolar damping has a reasonable chance of changing the turbulent
power spectrum and consequently the stochastic reconnection speed.
The ISM is extremely heterogeneous, but we can illustrate the 
effects of turbulence using the same idealized
phases from \citet{DL98}.  We have evaluated the reconnection speed
using equations (\ref{als}) and (\ref{lim2b}) for each phase.
Our results are given in the second to last line of Table~1, and we give a detailed
discussion below.  (The last line includes the effects of tearing modes,
which appear to dominate in the densest parts of molecular clouds). 
The turbulence which creates
field line stochasticity is assumed to be supplied at some large
scale, specified below, so we are implicitly ignoring the possibility
that as reconnection proceeds it will provide a local source of
turbulence\footnote{The increase of stochasticity due to reconnection
may be the source of the finite time instability associated with
solar flares \citep{lv99,VL00}.}
We have also ignored the possibility that reconnection can
change the ionization balance near the current sheet.  

We see from an inspection of Table~1 that neutral drag produces
a limit on the reconnection speed which is usually more restrictive than
the limit given at $L'=L$, that is, the limit which applies
to an ionized plasma. The only exception is the warm ionized
medium, where the neutral content is $\sim 1$\% and the turbulent
cascade continues down to $r_L$.  Assuming
strong turbulence, with a Mach number of order unity, the reduction
in the reconnection speed from the ionized limit is only a factor
of $\sim10$ for neutral atomic gas.  This is moderately sensitive to the scale
of turbulent energy injection, but since we expect that $L\sim l$
this will not be large effect.
On the other hand, strongly subsonic turbulence
implies a much smaller limit on $V_{rec}$.  Quiescent regions
in the ISM should have reconnection speeds much smaller than the
local Alfv\'en speed.

Molecular gas occupies only a small fraction of the ISM's volume,
but plays a special role as the site of star formation.  Table~1
indicates that as we consider increasingly dense molecular gas,
we move towards the case of decoupling above the viscous damping
scale.  In this limit the reconnection speed becomes more sensitive to
the energy injection scale.  We have assumed a single (large)
scale here, but it may be that a much smaller value is appropriate
in small dense regions.  In that case the limit on the reconnection
speed would drop significantly.  However,
if the density increases to the point where resistive instabilities
produce current sheets substantially broader than the width
given in equation (\ref{dmin1}) then the limit on the reconnection
speed will once again increase.  This accounts for the double
entry for molecular clouds.  Following the
approach used in other phases we get that
the reconnection speed is lower than
the Alfv\'en speed, even for a Mach number of 1, by more than
an order magnitude.
Unlike the atomic phases, this result scales as the square root of the
turbulent velocity, so that weaker turbulence has a less dramatic effect on the
reconnection speed. 
However, this result can be a substantial underestimate.  A large scale current
sheet is unstable to tearing modes (\cite{FKR63}), and these provide a localized
source of turbulence which will broaden the current sheet.  This sets a minimum
value of the current sheet width.  The resulting limit on $V_{rec}$ takes
precedence when it gives a larger $\Delta$, and a larger $V_{rec}$, than
estimates based purely on the turbulent cascade.

We show the effect of tearing modes in the last line of Table~1.  
For $\aleph_c<1$, for example, in the dense cores of molecular clouds, we can
take the parallel correlation scale as
\be
k_{\|}'={1\over V_A}{\rho_i\over\rho t_{in}}.
\ee
The corresponding current sheet width, set by tearing mode instabilities is
\be
\Delta\sim {1\over k_{\|}'}\left({\eta\rho_i\over t_{in}\rho V_A^2}\right)^{3/10},
\ee
for the width of the broadened current sheet.  This result is based on the
tearing mode dispersion relation for a current sheet embedded in an
neutral substrate \citep{Z89}, and is equation (\ref{app})
in the appendix.
For motion over scales comparable to $k'^{-1}_{\|}$ we have  
$V_{eject}\sim V_A$ so equation (\ref{v2})
implies
\be
V_{rec}\le V_A k_{\|}'L
\left({\eta\rho_i\over t_{in}\rho V_A^2}\right)^{3/10}
\approx V_A\left({\eta\over LV_A}\right)^{3/10}
\left({\rho_i\over\rho}{L\over V_At_{in}}\right)^{13/10}
=V_A\left({V_AL\over\eta_{ambi}}\right)\left({\eta\over\eta_{ambi}}\right)^{3/10},
\ee
where $\eta_{ambi}$ is the ambipolar diffusion coefficient defined by
\be
\eta_{ambi}\equiv {\rho\over\rho_i} V_A^2 t_{in},
\ee
and $\eta<\eta_{ambi}<V_AL$.

In a dense neutral medium the electron collision
rate is dominated by collisions with neutrals, for which \citet{DRD83}
give
\be
t_{en}^{-1}\approx 2.62\times10^{-9} n_n\left({T_e\over 10 K}\right)^{1/2}.
\ee
Consequently the resistivity for DC phase of the ISM is approximately
$9.3\times10^9$ cm$^2$ sec$^{-1}$.  Combining these results we obtain the
last entry of Table~1.  The large coefficient implies that the usual large
scale limit, given in equation (\ref{eq:lim2a}), will usually apply here. 
In the densest and most neutral phases of the ISM we recover the fast reconnection
speeds typical of an ionized plasma.

For $\aleph_c>1$ we need to consider tearing modes in a viscous medium, with a parallel
wavenumber $\sim k_{\|,c}$.  In this case we find 
\be
\Delta\sim k_{\|,c}^{-1}\left({\eta k_{\|,c}\over V_A}\right)^{3/16}
\left({\nu k_{\|,c}\over V_A}\right)^{3/16},
\ee
which is equation (\ref{app2}) in the appendix.  Using this expression, and remembering that
$k_c^2\nu=k_{\|,c}V_A$, we obtain
\be
V_{rec}\le V_A (k_{\|,c}L)\left({k_c\over k_{\|,c}}\right)^{7/8}\left({\eta k_{\|,c}\over V_A}\right)^{1/8}.
\ee
The resistivity for our idealized molecular cloud phase is $\sim 1.3\times10^8$.  The
resultant value of the reconnection speed in molecular clouds is competitive with the
limit derived from turbulence, but is slightly smaller (meaning that the larger limit
applies).  Evidently tearing modes become rapidly more important as we go to 
the densest regions of molecular clouds, and are unimportant outside of such regions.



Finally, we note that for collisional fluids the magnetic Prandtl number,
$\nu/\eta$, is typically very small, so that the turbulent
cascade extends down to resistive scales.  In this case
equation (\ref{eq:lim2a}) is our best estimate of the reconnection
speed, assuming, as before, that $v_l\le V_A$.  As an example we can
consider the temperature minimum in the solar photosphere. Adopting
$T\approx 4200$, $n_e\approx 10^{11}$ cm$^{-3}$, and
$n_n\approx 10^{15.2}$ cm$^{-3}$, we find that
$\nu\sim c_n^2t_n\approx 2\times10^5$cm$^2$sec$^{-1}$, and
$\eta\sim 3\times10^9$cm$^2$sec$^{-1}$.

\section{Conclusions}

Our results can be summarized as follows:

\begin{enumerate}
\item
In plasmas characterized by a moderate magnetic Prandtl number
(resistivity less than, or of order, viscosity) magnetic reconnection follows
the model described in LV99, so that reconnection
is fast in the presence of noise.  Collisionless plasmas
with small neutral fractions and collisionally dominated
plasmas are typically in this class.   In either case
the turbulent cascade proceeds to either the resistive scale or
the ion Larmor radius.
For typical interstellar conditions, it is the ion Larmor radius
which is larger.

\item
As the neutral content
in plasma increases the turbulent cascade is interrupted at a scale greater than
the neutral mean free path. However, the magnetic field perturbations
are not suppressed below this scale.  On the contrary, we find that the
magnetic field power spectrum flattens out, and exhibits more power on
small scales than the Goldreich-Sridhar spectrum. Velocity fluctuations
in this regime are driven by magnetic field perturbations, whose energy is much
larger than the kinetic energy perturbations on the same scale.
The importance of this
new regime of magnetic stochasticity goes far beyond understanding the reconnection
problem.

\item
On somewhat smaller scales, once the ions and neutrals are
decoupled, the turbulent cascade can reappear.  The warm atomic gas
in the ISM shows only a slight interruption in the turbulent
cascade and the small scale scintillations are close to a continuation
of the large scale power.  However, this break will be more conspicuous
in colder and denser phases of the ISM.  In this limit the small
scale turbulence will be strongly intermittent.

\item
Magnetic stochasticity on scales smaller than the ion-neutral damping
cut-off of the Goldreich-Sridhar spectrum promotes fast reconnection.
The diffusion of magnetic field lines initially decreases rapidly below this
scale, but the decoupling of ions and neutrals leads to a minimum value
of the diffusion coefficient on small scales.  The most stringent limit on
reconnection speeds comes from considering the largest scales where this minimum value
holds.   This is in the range of the turbulent cascade where ions and
neutrals are decoupled, but neutral drag is still sufficient to stabilize
magnetic perturbations. In most phases of the ISM this gives a reconnection speed
which is just slightly smaller than the ionized plasma estimate
given in LV99.

\item
Our study shows that the reconnection speed in the ISM is
always much faster than the predictions given by the Sweet-Parker rate.
Even in very dense and cold regions it is only slightly below the
Alfv\'en speed, and tearing mode instabilities in the current sheet tend
to drive it back up to the Alfv\'en speed.
This process needs to be considered
in models for magnetic flux removal
during star formation, and for studies of the dynamics of magnetized molecular
clouds.
\end{enumerate}

Finally, we note that our conclusions are moderately sensitive to the
nature of field line diffusivity in the various damped regimes
($k_t<k<\min[k',k_c]$).  We have argued that the shearing due to
motions at $k\sim k_c$ dominate until we are well into the decoupled
damped regime.  Our estimate for a constant field line diffusivity
on somewhat smaller scales has not been tested against simulations.
If this turns out to be too high (or low) then our reconnection
speed estimate will have to be lowered (or raised).

\acknowledgements

AL and JC acknowledge NSF grant AST-0125544 and ETV acknowledges
the support of the NSF grant AST-0098615.  ETV would also like to
thank the hospitality of the organizers of the Festival de Theorie
in Aix-le-Provence, 2001, where some of this work was done.

\appendix
\section{Tearing Modes and Current Sheet Width}

The current sheet implicit in the Sweet-Parker picture of magnetic
reconnection is well known to be unstable to tearing modes
\citep{FKR63}, which will create a turbulent zone between the
volumes of unreconnected magnetic flux.
In LV99 we suggested that the broadening of the current
sheet implied by this effect should lead to an enhanced reconnection
rate for laminar field lines, on the order of
\be
V_{rec}\approx V_A\left({\eta\over V_A L}\right)^{3/10}.
\ee
Broadly similar results were obtained by \citet{S88}, except that we
have calculated the typical size of the magnetic field fluctuations 
using tearing mode theory rather than leaving it as a free parameter.

\citet{Z89} derived
the dispersion relation for tearing modes in
a current sheet of thickness $\Delta$ embedded in a neutral substrate.
It is
\be
{1\over (k_{\|}\Delta^2)^4}={\gamma^5\rho_i\over k_{\|}^2V_A^2\rho}
{\Delta^2\over\eta^3}\left(1+{t_{in}^{-1}\over\gamma+(\rho_i/\rho)t_{in}^{-1}}\right),
\label{a1}
\ee
where $\gamma$ is the tearing mode growth rate and 
$k_{\|}\Delta<<1$.  This favors modes with minimal $k_{\|}$, i.e. transverse
wavelengths comparable to the current sheet length.  The resulting
turbulence will widen the current sheet until the tearing modes are marginally
stabilized by the shear due to the ejection of plasma, i.e. 
$\gamma\approx V_A/L\approx k_{\|}V_A$.

When the ions and neutrals are decoupled the ejection velocity from the current
sheet is
\be
V_{eject}\approx k_{\|}V_A^2{\rho\over\rho_i}t_{in}.
\ee
The dispersion relation for $\gamma\ge (\rho_i/\rho)t_{in}^{-1}$ is
\be
\gamma^4={V_A^2\eta^3 t_{in}\over \Delta^{10} k_{\|}^2}.
\ee
So for the saturated modes
\be
\Delta k_{\|}\approx \left({\eta\rho_i\over V_A^2\rho t_{in}}\right)^{3/10}.
\label{app}
\ee

This determines the width of the outflow zone if we can assume that
the tearing modes produce stochastic mixing of the field lines, that is, that 
neighboring field lines are well-mixed within this volume.  In two dimensions
this is obviously not the case.  In three dimensions it is probably valid.
We are also assuming that the existence of turbulence in the current sheet
does not lead to widespread mixing of field lines outside the current sheet.
Although some mixing seems inevitable, the fact that the modes are driven
only as long as they are narrow enough to fit within the current sheet,
and grow only about as fast as plasma, and magnetic structures, are ejected
from the current sheet, suggests that any induced field line mixing will
only spread the the outflow zone by an additional factor of order unity. 

When the ions and neutrals are {\it not} decoupled, but viscous drag
plays an important role we can rewrite equation (\ref{a1}) as
\be
\left({\Delta'\over I}\right)^4={\gamma^5\rho_i\over k_{\|}^2V_A^2\rho}
{\Delta^2\over\eta^3}\left(1+{(\nu/\Delta^2)\over\gamma}\right),
\ee
where we have assumed a single fluid ($\gamma> (\rho_i/\rho)t_{in}^{-1}$)
and replaced drag by a neutral substrate with viscous damping.
The ejection velocity in this case is obtained by balancing viscous drag
with magnetic forces,
\be
k_{\|}V_A^2\sim {\nu\over \Delta^2}V_{eject}.
\ee
We conclude that in this case
\be
\Delta\sim k_{\|,c}^{-1}\left({\eta k_{\|,c}\over V_A}\right)^{3/16}
\left({\nu k_{\|,c}\over V_A}\right)^{3/16}.
\label{app2}
\ee

%
%

\clearpage

\begin{deluxetable}{lccccc}
\tablecolumns{6}
\rotate
\setlength{\tabcolsep}{0.02in}
\tabletypesize{\scriptsize}
\tablecaption{Scales and Reconnection speeds for idealized phases of the ISM}
\tablehead{
\colhead{ISM:}      & \colhead{WIM} &
\colhead{WNM}      & \colhead{CNM}       &
\colhead{MC}      & \colhead{DC}}
\startdata
$n$&$0.1$&$0.4$&$30$&$300$&$10^4$\\[0.4cm]
$x$&$0.99$&$0.1$&$10^{-3}$&$10^{-4}$&$10^{-6}$\\[0.4cm]
$T$&$8000$&$6000$&$100$&$20$&$10$\\[0.4cm]
$F_M$&$0.14$&$0.43$&$0.43$&\nodata&\nodata\\[0.4cm]
$\aleph_c$&$9.1l_{30}^{1/2}{\cal M}^{-1}\beta^{1/4}$&$16l_{30}^{1/2}{\cal M}^{-1}\beta^{1/4}$
&$7.8l_{30}^{1/2}{\cal M}^{-1}\beta^{1/4}$
&$4.8l_{30}^{1/2}{\cal M}^{-1}\beta^{1/4}$&$0.37l_{30}^{1/2}{\cal M}^{-1}\beta^{1/4}$\\[0.4cm]
$k_c^{-1}$&\nodata&$2.4\times10^{16}l_{30}^{1/4}{\cal M}^{-1/2}\beta^{-1/8}$
&$3.4\times10^{15}l_{30}^{1/4}{\cal M}^{-1/2}\beta^{-1/8}$
&$8.8\times10^{13}l_{30}^{1/4}{\cal M}^{-1/2}\beta^{-1/8}$
&\nodata\\[0.4cm]
\tableline
$k_{\|,c}^{-1}$&\nodata&$3.8\times10^{17}l_{30}^{1/2}{\cal M}^{-1}\beta^{-1/4}$
&$1.0\times10^{17}l_{30}^{1/2}{\cal M}^{-1}\beta^{-1/4}$
&$9.0\times10^{15}l_{30}^{1/2}{\cal M}^{-1}\beta^{-1/4}$
&\nodata
\\[0.4cm]
$k'^{-1}$&\nodata&\nodata&\nodata&\nodata&$2.6\times10^{13}l_{30}^{-1/2}{\cal M}\beta^{-1/2}$\\
[0.4cm]
$k_{\|}^{\prime -1}$&\nodata&\nodata&\nodata&\nodata
&$4.0\times10^{15}\beta^{-1/2}$\\
[0.4cm]
$k_p^{-1}$&\nodata&$9.6\times10^{15}l_{30}^{1/12}{\cal M}^{-1/6}\beta^{-5/24}$
&$1.7\times10^{15}l_{30}^{1/12}{\cal M}^{-1/6}\beta^{-5/24}$
&$5.2\times10^{13}l_{30}^{1/12}{\cal M}^{-1/6}\beta^{-5/24}$
&\nodata
\\[0.4cm]
$k_d^{-1}$&\nodata&$6.4\times10^{15}\beta^{-1/4}$&$1.2\times10^{15}\beta^{-1/4}$
&$4.0\times10^{13}\beta^{-1/4}$&\nodata\\[0.4cm]
\tableline
$k_t^{-1}$&\nodata&$1.7\times10^{14}{\cal M}^{1/2}l_{30}^{-1/4}\beta^{-3/4}$&
$1.3\times10^{13}{\cal M}^{1/2}l_{30}^{-1/4}\beta^{-3/4}$&
$1.8\times10^9{\cal M}^{1/2}l_{30}^{-1/4}\beta^{-3/4}$&
$2.6\times10^7{\cal M}l_{30}^{-1/2}\beta^{-1/2}$
\\[0.4cm]
$k_{\|,t}^{-1}$&\nodata&
$8.5\times10^{15}\beta^{-1/2}$
&$1.3\times10^{14}\beta^{-1/2}$&
$1.9\times10^{13}\beta^{-1/2}$
&$4.0\times10^{15}\beta^{-1/2}$\\[0.4cm]
$r_L$&$7.3\times10^7\beta^{1/2}$&$1.2\times10^8\beta^{1/2}$&
$1.3\times10^{8}\beta^{1/2}$&$1.3\times10^8\beta^{1/2}$&
$2.3\times10^8\beta^{1/2}$\\[0.4cm]
$V_{rec}$&$v_T{L_{30}^{1/2}\over l_{30}^{1/2}}\min[1,l_{30}/L_{30}]$&
$0.092v_T{\cal M}^2\beta^{-1/2}{L_{30}\over l_{30}^{3/2}}$&
$0.077v_T{\cal M}^2\beta^{-1/2}{L_{30}\over l_{30}^{3/2}}$&
$0.095v_T{\cal M}^2\beta^{-1/2}{L_{30}\over l_{30}^{3/2}}$&
$0.094v_T{\cal M}^{-1/2}\beta^{3/4}{L_{30}\over l_{30}^{1/4}}$\\[0.4cm]
$V_{rec}(TM)$&\nodata&\nodata&\nodata&$0.05v_T{\cal M}^{-17/8}l^{9/16}_{30}L_{30}\beta^{-61/64}$&
$21v_T{\cal M}^{-1}\beta^{-0.3}L_{30}$
\enddata
\tablecomments{Here ${\cal M}$ is the turbulence Mach number,
$\equiv v_T/c_n$, $\beta\equiv c_n^2/V_A^2$, $F_M$ is the (approximate) 
mass fraction contained in each phase,
and a subscript `$30$' indicates that distances are given in
units of $30$ parsecs. The second row gives the
ionization fraction for each phase.  The abbreviations in the first row
denote `Warm Ionized Medium', `Warm Neutral Medium', `Cold Neutral Medium',
`Molecular Cloud', and `Dense Core' (in a molecular
cloud). Length scales are defined in the text and are given here in
centimeters.}
\end{deluxetable}
\clearpage

\begin{figure}
\plotone{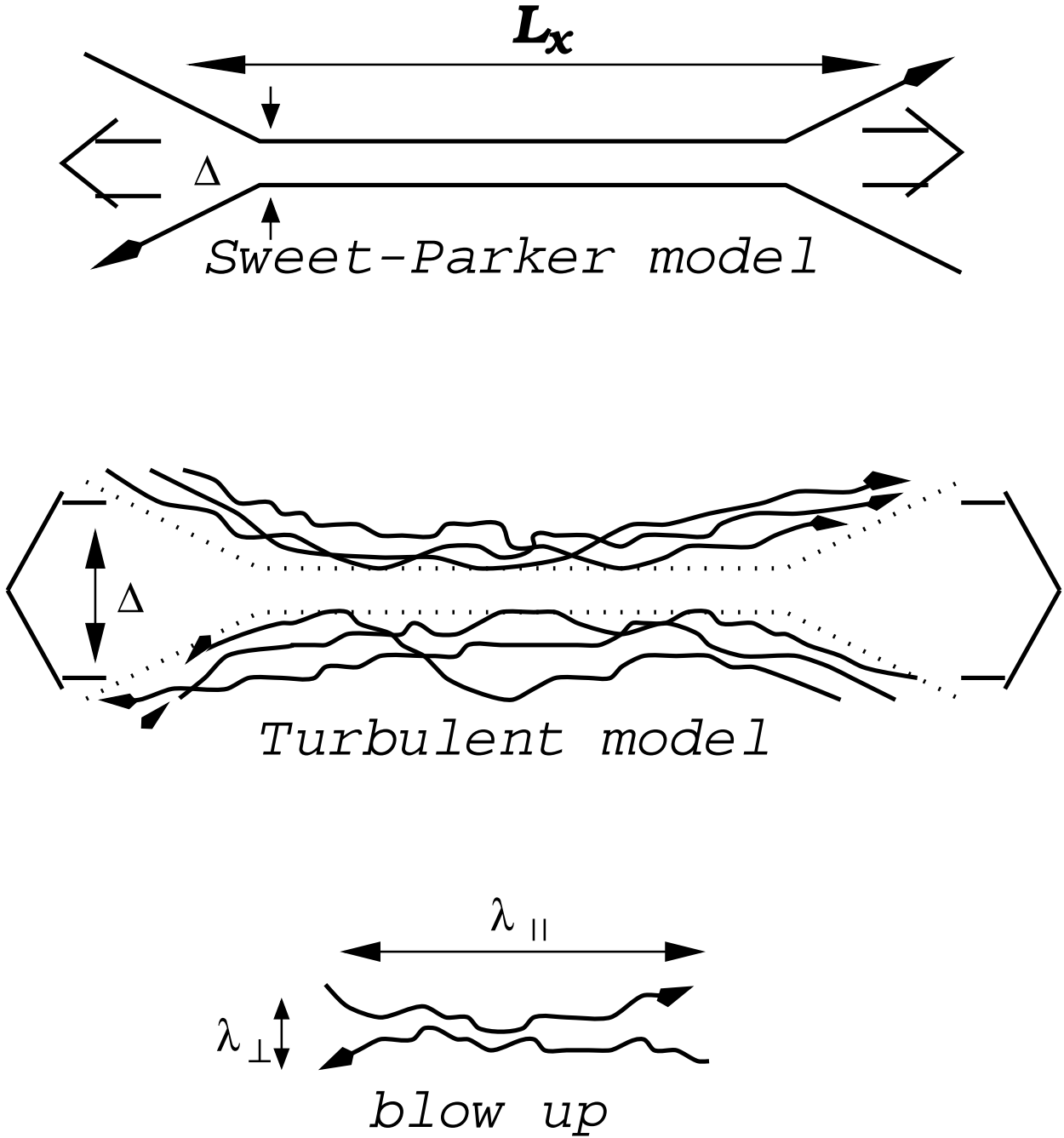}
\caption{Upper plot: Sweet-Parker scheme of reconnection. Middle plot:
illustration of stochastic reconnection that accounts for field line
noise.  Lower plot: a close-up of the contact region.
Thick arrows depict outflows of plasma. From \cite{lv00}.}
\end{figure}

\clearpage

\begin{figure}
\plotone{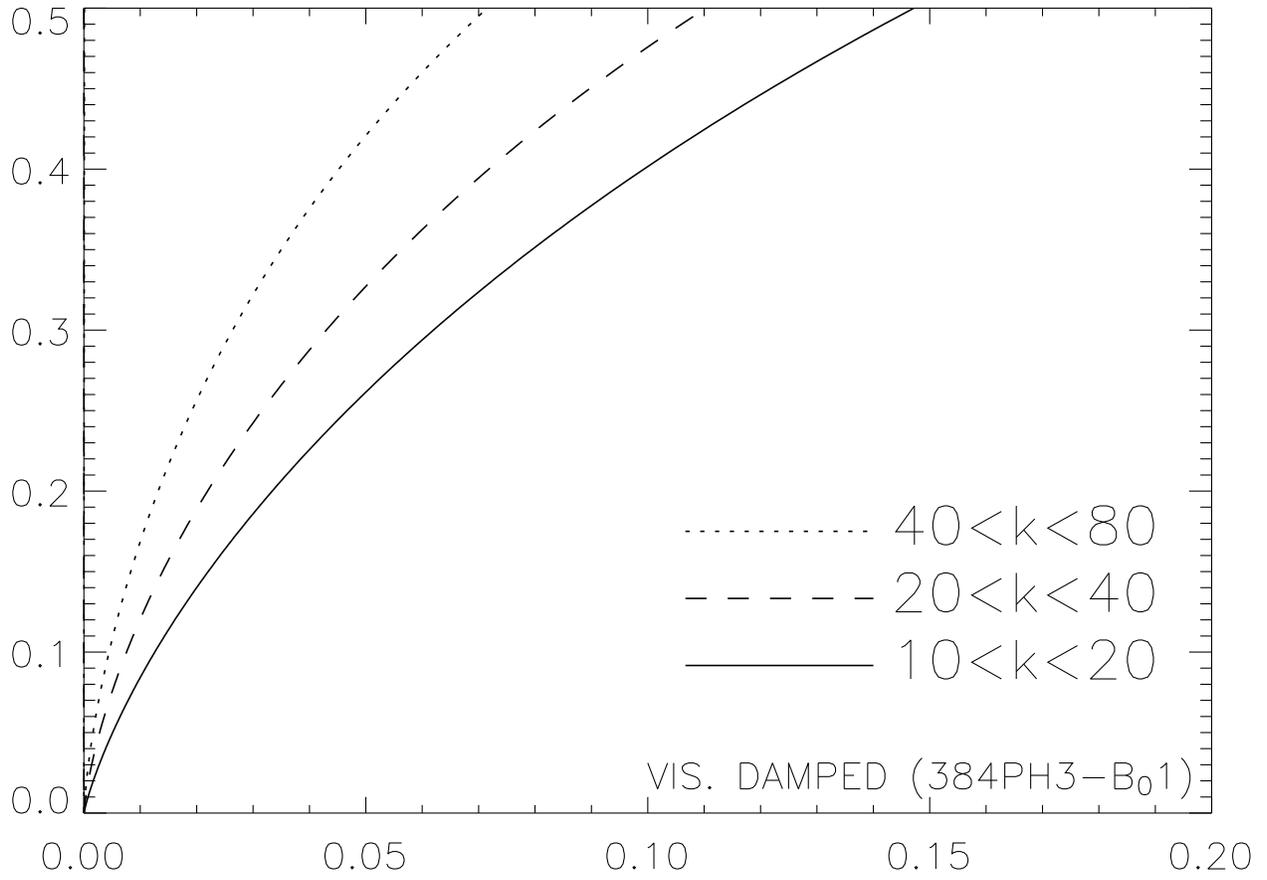}
\caption{Fractional volume (X-axis) vs.
      fractional magnetic energy in the volume (Y-axis)
for viscosity-damped MHD turbulence:
        smaller scales show a higher concentration of magnetic energy.
}
\end{figure}

\clearpage

\begin{figure}
\epsscale{.60}
\plotone{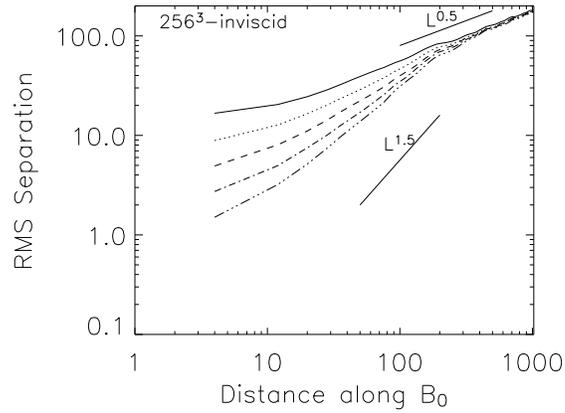}
\caption{Root mean square separation of field lines in a simulation of
inviscid MHD turbulence, as a function of
distance parallel to the mean magnetic field, for a range of initial
separations.  Each curve represents 1600 line pairs.
The simulation has been filtered to remove pseudo-Alfv\'en
modes, which introduce noise into the diffusion calculation.}
\end{figure}

\clearpage

\begin{figure}
\epsscale{.40}
\plotone{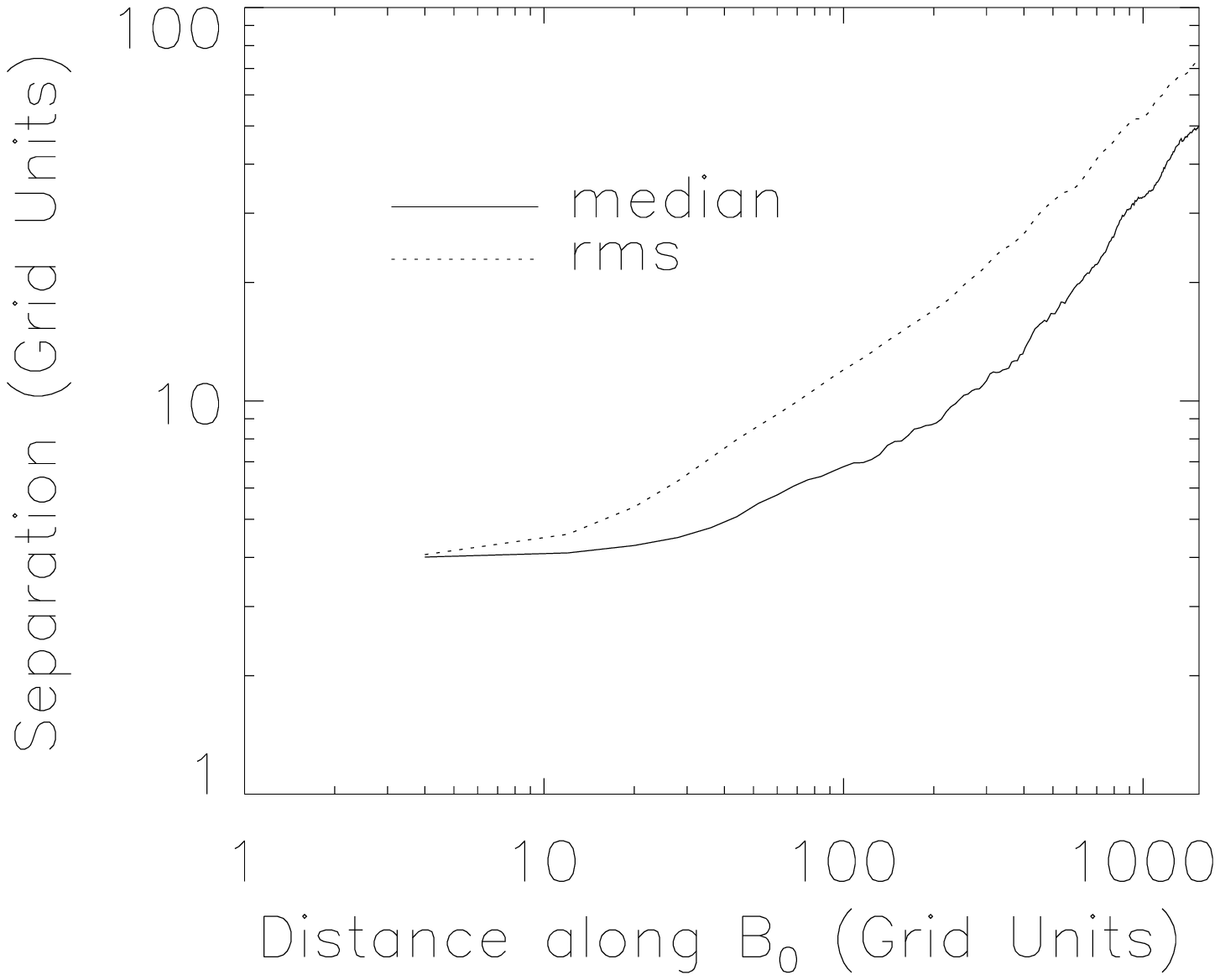}
\plotone{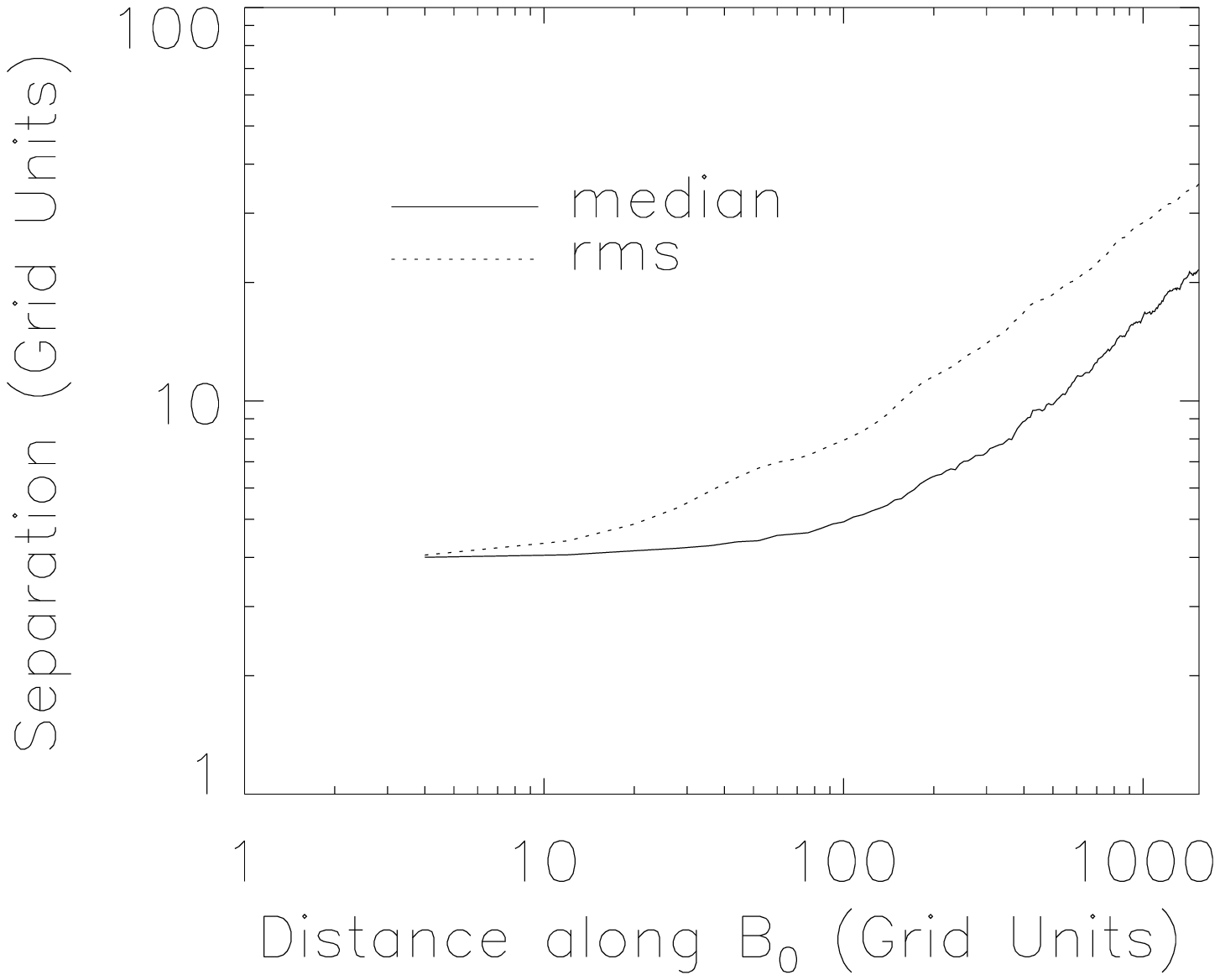}
\plotone{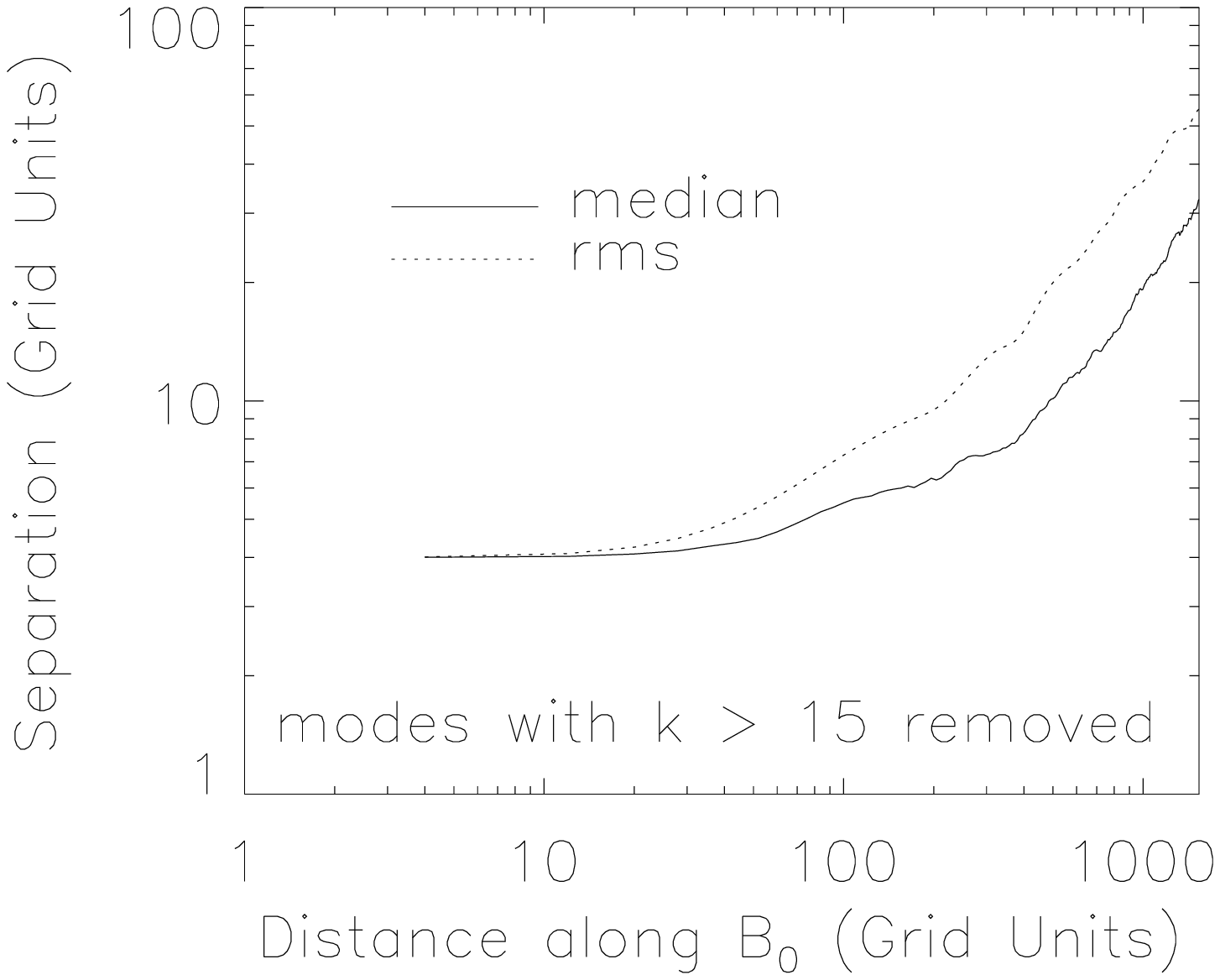}
\caption{Root mean square and median separation of field lines in simulation of
viscously damped MHD turbulence, as a function of distance parallel to
the mean magnetic field.  In (a) we see the total result.  In (b) we have
removed long wavelength modes.  In (c) we retain only the long wavelength
modes.  In all three cases the median tracks the mean, at a slightly
smaller amplitude, and is consistent with exponential growth. The simulation
used a $384^3$ grid.  The data are based on 1600 line pairs.}
\end{figure}

\clearpage

\begin{figure}
\plotone{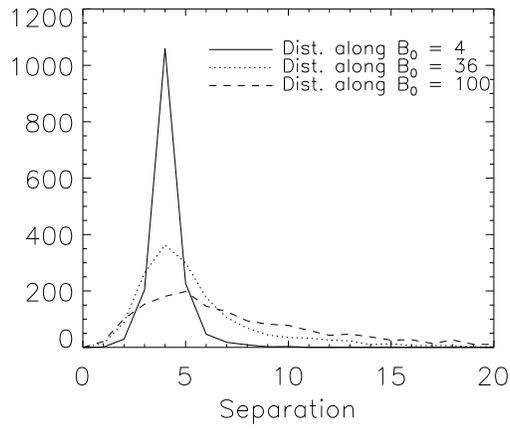}
\caption{A histogram of field line separations for viscously damped
turbulence.  The distribution shown here is the same one used for the
curve in figure 4a for an initial separation of 4 grid points.
The tail of large separations grows smoothly and
continuously, and involves a large fraction of all field lines despite
the intermittency of the magnetic field perturbations.}
\end{figure}

\begin{thebibliography}{}
\bibitem[Armstrong, Rickett \& Spangler(1997)]{ARS95}
Armstrong, J.W., Rickett, B.J. \& Spangler, S.R. 1995,\apj, 443, 209
\bibitem[Bhattacharjee \& Hameiri(1986)]{BH86}Bhattacharjee, A. \&
Hameiri, E.\ 1986, \prl, 57, 206
\bibitem[Bhattacharjee, Ma, \& Wang(2001)]
{BMW01} Bhattacharjee, A., Ma, Z.W. \& Wang, X. 2001, Phys. Plasmas, 8(5), 1829
\bibitem[Bhattacharjee, Ma, \& Wang(2003)]
{BMW03} Bhattacharjee, A., Ma, Z.W. \& Wang, X. 2003,
in {\it Turbulence and Magnetic Fields in Astrophysics}, 
                eds. T. Passot \& E.
                Falgarone (Springer Lecture Notes in Physics 614; 2003), 
	351
\bibitem[Biskamp(1996)]{B96}Biskamp, D.\ 1996, Astrophys. \& Sp. Sci., 242,165
\bibitem[Biskamp(2000)]{B00}Biskamp, D.\ 2000, Magnetic Reconnection
in Plasmas (Cambridge: Cambridge University Press)
\bibitem[Biskamp, Schwarz \& Drake(1997)]{BSD97}
Biskamp, D., Schwarz, E. \& Drake, J.F. 1997, Phys. Plasmas,
4, 1002
\bibitem[Brandenburg(2001)]{B01}Brandenburg, A.\ 2001, \apj, 550, 824
\bibitem[Cattaneo \& Hughes(1996)]{CH96}Cattaneo, F., \& Hughes, D.W.\ 1996,
\pre, 54, 4532
\bibitem[Chandran \& Maron(2003)]{CM03}Chandran, B.D.G. \& Maron, J.L.
\ 2003, astro-ph/0303214, submitted to \prl
\bibitem[Cho \& Lazarian(2002)]{CL02} 
Cho, J. \& Lazarian A. 2002, Phys. Rev. Lett., 88, 245001
\bibitem[Cho \& Lazarian(2003)]{CL03} 
Cho, J. \& Lazarian A. 2003, MNRAS, 345, 325
\bibitem[Cho \& Vishniac(2000)]{CV00}
Cho, J., \& Vishniac, E. T. 2000, \apj, 538, 217
\bibitem[Cho, Lazarian \& Vishniac(2002a)]{CLV02a}
Cho, J., Lazarian, A. \& Vishniac, E. T. 2002a, \apj, 564, 291
\bibitem[Cho, Lazarian \& Vishniac(2002b)]{CLV02b}
Cho, J., Lazarian, A. \& Vishniac, E. T. 2002b, \apj, 566, 49L
\bibitem[Cho, Lazarian \& Vishniac\ 2003a]{CLV03a}
Cho, J., Lazarian, A. \& Vishniac, E. T. 2003a, 
in {\it Turbulence and Magnetic Fields in Astrophysics}, 
                eds. T. Passot \& E.
                Falgarone (Springer Lecture Notes in Physics 614; 2003), 
	56
\bibitem[Cho, Lazarian \& Vishniac(2003b)]{CLV03b} Cho, J., Lazarian, A., \&
Vishniac, E.T. \ 2003b, \apj, 595, 812
\bibitem[Cho et al.(2003)]{CLHKKM03}Cho, J., Lazarian, A., Honein, A., 
Knaepen, B., Kassinos, S. \& Moin, P.\ 2003, \apj, 589L, 77
\bibitem[Dere(1996)]{D96} Dere, K.P. 1996, \apj, 472, 864
\bibitem[Draine, Roberge \& Dalgarno(1983)]{DRD83}Draine, B.T., Roberge,
W.G., \& Dalgarno, A. 1983, \apj, 264, 485
\bibitem[Draine \& Lazarian(1998)]{DL98}
Draine, B.T., \& Lazarian, A. 1998, \apj, 494, L19
\bibitem[Furth, Killeen, \& Rosenbluth(1963)]{FKR63} Furth, H.P.,
Killeen, J., \& Rosenbluth, M.N.\ 1963, Phys. Fluids, 6, 459
\bibitem[Goldreich \& Sridhar(1995)]{gs95} Goldreich, P. \& Sridhar, S.\ 1995,
\apj, 438, 763 (GS95)
\bibitem[Gruzinov \& Diamond(1994)]{GD94}Gruzinov, A.V. \& Diamond, P.H.\ 1994,
\prl, 72, 1651
\bibitem[Gruzinov \& Diamond(1996)]{GD96}Gruzinov, A.V. \& Diamond, P.H.\ 1996,
Phys. Plasmas, 3, 1853
\bibitem[Hameiri \& Bhattacharjee(1987)]{HB87}Bhattacharjee, A. \&
Hameiri, E.\ 1987, Phys. Fluids, 30, 1744
\bibitem[Heitsch \& Zweibel(2003a)]{HZ03a}Heitsch, F. \& Zweibel, E.G.\ 2003a,
\apj, 583, 229
\bibitem[Heitsch \& Zweibel(2003b)]{HZ03b}Heitsch, F. \& Zweibel, E.G.\ 2003b,
\apj, 590, 291
\bibitem[Hughes et al.(1996)]{HCK96}
Hughes, D.W., Cattaneo, F. \& Kim, E.J.\ 1996, Phys. Lett. A, 223, 167
\bibitem[Innes, Inhester, Axford \& Wilhelm(1997)]{IIAW97}
Innes, D.E., Inhester, B., Axford, W.I., \& Wilhelm, K. 1997, Nature, 386, 811
\bibitem[Jacobson \& Moses(1984)]{JM84}Jacobson, A.R. \& Moses, R.W. 1984,
Phys. Rev. A, 29(6), 3335
\bibitem[Ji, Yamada, Hsu \& Kulsrud(1998)]{JYHK98}Ji, H., Yamada, M., Hsu,
S. \& Kulsrud, R.\ 1998, \prl, 80, 3256 
\bibitem[Kim \& Diamond(2001)]{KD01} Kim, E.-J., \& Diamond, P.H.\ 2001,
\apj, 556, 1052
\bibitem[Krause \& Radler(1980)]{KR80} Krause, F., \& Radler, K.H. 1980,
Mean-Field Magnetohydrodynamics and Dynamo Theory (Oxford: Pergamon Press)
\bibitem[Kulsrud \& Pearce(1969)]{KP69} Kulsrud, R., \& Pearce, W.P. 1969,
\apj, 156,445
\bibitem[Lazarian \& Pogosyan(2000)]{LP00} 
Lazarian, A. \& Pogosyan, D.\ 2000, \apj, 537, 720
\bibitem[Lazarian \& Vishniac(1999)]{lv99}
Lazarian, A. \& Vishniac, E.T. 1999, \apj, 517, 700 (LV99)
\bibitem[Lazarian \& Vishniac(2000)]{lv00} Lazarian, A. \& Vishniac, E.T.
\ 2000, Rev.Mex.~de~Astron.~y~Astrof., 9, 55
\bibitem[Lithwick \& Goldreich(2001)]{LG01}Lithwick, Y. \& Goldreich, P.
\ 2001, \apj, 562, 279
\bibitem[Maron \& Goldreich(2001)]{MG01}
Maron, J. \& Goldreich, P. 2001, \apj, 554, 1175
\bibitem[Maron, Chandran \& Blackman(2003)]{MCB03}Maron, J.L., Chandran,
B.D.G. \& Blackman, E.G.\ 2003, astro-ph/0303217
\bibitem[Matthaeus \& Lamkin(1985)]{ML85} Matthaeus, W.H. \& Lamkin, S.L.
\ 1985, Phys. Fluids, 28, 303
\bibitem[McIvor(1977)]{M77}McIvor, I. 1977, \mnras, 178, 85
\bibitem[Minter \& Spangler(1997)]{MS97}Minter, A.H., \& Spangler, S.R.\ 1997,
\apj, 485, 182
\bibitem[Moffatt(1978)]{M78} Moffatt, H.K. 1978, Magnetic Field Generation in E
lectrically
Conducting Fluids (Cambridge: Cambridge University Press)
\bibitem[Naidu, McKenzie \& Axford(1992)]{NMA92}
Naidu, K., McKenzie, J.F., \& Axford, W.I. 1992, Ann.
Geophys., 10, 827
\bibitem[Narayan \& Medvedev (2001)]{MN01}
Narayan, R. \& Medvedev, M.V.\ 2001, \apj, 562L, 129
\bibitem[Parker(1957)]{P57}Parker, E.N.\ 1957, J. Geophys. Res., 62, 509
\bibitem[Parker(1979)]{P79}\rule{1.2cm}{0.2mm}\ 1979, Cosmical
Magnetic Fields (Oxford: Clarendon Press)
\bibitem[Parker(1992)]{P92}Parker, E.N.\ 1992, \apj, 401, 137
\bibitem[Petschek(1964)]{P64}Petschek, H.E.\ 1964,
{\it The Physics of Solar Flares}, AAS-NASA
Symposium, NASA SP-50 (ed. W.H. Hess), Greenbelt, Maryland, p.~425
\bibitem[Priest \& Forbes(2000)]{PF00}Priest, E. \& Forbes, T.\ 2000,
Magnetic Reconnection: MHD Theory and Applications (Cambridge: Cambridge
University Press)
\bibitem[Schekochihin, Maron, Cowley \& McWilliams(2002)]{SMCM02}
Schekochihin, A., Maron, J., Cowley, S. \& McWilliams, J.\ 2002, \apj,
576, 806
\bibitem[Schekochihin, Cowley, Maron \& McWilliams(2003)]{SCMM03}
Schekochihin, A., Cowley, S., Maron, J. \& McWilliams, J.
\ 2003, astro-ph/0308336 
\bibitem[Shay \& Drake(1998)]{SD98}
Shay, M.A., Drake, J.F. 1998, Geophys. Res. Let., 25(20), 3759
\bibitem[Shay, Drake, Denton \& Biskamp(1998)]{SDD98}
Shay, M.A., Drake, J.F., Denton, R.E., \& Biskamp, D.
1998, J. Geophys. Res., 103, 9165
\bibitem[Spangler(1991)]{S91} Spangler,S.R. \ 1991,\apj, 376, 540
\bibitem[Spangler(1999)]{Sp99} Spangler,S.R. \ 1999,\apj, 522, 879
\bibitem[Speiser(1970)]{S70}Speiser, T.W. 1970, Planet. Space Sci., 18,
613
\bibitem[Spitzer(1978)]{S78} Spitzer, L.\ 1978, Physical Processes in the
Interstellar Medium (New York: John Wiley \& Sons)
\bibitem[Stanimirovic \& Lazarian(2001)]{SL01}Stanimirovic, S \& Lazarian, A.\ 2001, \apj, 441, 53
\bibitem[Strauss(1985)]{S85} Strauss, H.R.\ 1985, Phys. Fluids, 28, 2786
\bibitem[Strauss(1988)]{S88} Strauss, H.R.\ 1988, \apj, 326, 412 
\bibitem[Sweet(1958)]{S58} Sweet, P.A.\ 1958, in IAU Symp. 6, Electromagnetic
Phenomena in
Cosmical Plasma, ed. B. Lehnert (New York: Cambridge Univ. Press), 123
\bibitem[Trintchouk, Yamada, Ji, Kulsrud \& Carter(2003)]{TYJKC03}
Trintchouk, F., Yamada, M., Ji, H., Kulsrud, R.M. \& Carter, T.A.\ 2003,
Physics of Plasmas, 10(1), 319
\bibitem[Vainshtein \& Cattaneo(1992)]{VC92}Vainshtein, S.I. and Cattaneo, F.
\ 1992, \apj, 393, 165
\bibitem[Vishniac \& Lazarian(1999)]{VL99}
Vishniac, E.T. \& Lazarian, A.\ 1999, \apj, 511, 193
\bibitem[Vishniac \& Lazarian(2000)]{VL00} 
Vishniac, E.T., \& Lazarian, A. 2000, in 
{\it Plasma Turbulence and Energetic Particles},
                ed. by M. Ostrowski, R. Schlickeiser  
                (Cracow, 2000) p.182
\bibitem[Vishniac \& Cho(2001)]{VC01}
Vishniac, E.T. \& Cho, J.\ 2001, \apj, 550, 752
\bibitem[Yan \& Lazarian(2002)]{YL02} 
Yan, H. \& Lazarian, A. 2002, Phys. Rev. Lett, 89, 281102
\bibitem[Zweibel(1989)]{Z89}Zweibel, E.G.\ 1989, \apj, 340, 550
\bibitem[Zweibel \& Brandenburg(1997)]{ZB97}
Zweibel, E.G., \& Brandenburg, A., 1997, \apj, 478, 563
\end{thebibliography}
\end{document}